\documentclass[sigconf]{acmart}
\setcopyright{none}
\renewcommand\footnotetextcopyrightpermission[1]{}

\usepackage{tikz}
\usepackage{amsmath}
\usepackage{xurl}
\usepackage{tablefootnote}
\usepackage{multirow}
\usepackage{placeins}
\usepackage{subcaption}

\usepackage{xcolor}
\usepackage[linesnumbered,ruled,vlined]{algorithm2e}
\newtheorem*{thm}{Game}
\usepackage{balance}

\begin{document}
\title{QueryCheetah: Fast Automated Discovery of Attribute Inference Attacks Against Query-Based Systems}

\author{Bozhidar Stevanoski}
\affiliation{%
  \institution{Imperial College London}
  \city{London}
  \country{United Kingdom}}
\email{b.stevanoski@imperial.ac.uk}

\author{Ana-Maria Cretu}
\authornote{Ana-Maria Cretu did most of her work while at Imperial College London.}
\affiliation{%
  \institution{EPFL}
  \city{Lausanne}
  \country{Switzerland}}
\email{ana-maria.cretu@epfl.ch}

\author{Yves-Alexandre de Montjoye}
\affiliation{%
  \institution{Imperial College London}
  \city{London}
  \country{United Kingdom}}
\email{demontjoye@imperial.ac.uk}

\begin{abstract}
  Query-based systems (QBSs) are one of the key approaches for sharing data. QBSs allow analysts to request aggregate information from a private protected dataset. Attacks are a crucial part of ensuring QBSs are truly privacy-preserving. The development and testing of attacks is however very labor-intensive and unable to cope with the increasing complexity of systems. Automated approaches have been shown to be promising but are currently extremely computationally intensive, limiting their applicability in practice. We here propose QueryCheetah, a fast and effective method for automated discovery of privacy attacks against QBSs. We instantiate QueryCheetah on attribute inference attacks and show it to discover stronger attacks than previous methods while being 18 times faster than the state-of-the-art automated approach. We then show how QueryCheetah allows system developers to thoroughly evaluate the privacy risk, including for various attacker strengths and target individuals. We finally show how QueryCheetah can be used out-of-the-box to find attacks in larger syntaxes and workarounds around ad-hoc defenses.
  \footnote{This is an extended version of the ACM CCS paper (\url{https://doi.org/10.1145/3658644.3690272}), which includes appendices.}
  \footnote{The code for this paper is available at \url{https://github.com/computationalprivacy/querycheetah}.}
\end{abstract}

\maketitle

\section{Introduction}
In the era of digital connectivity, we are generating data on an unprecedented level~\cite{sagiroglu2013big}. These data are collected on a large scale, which opens the door to new applications such as real-time traffic congestion update systems~\cite{blogGoogleMaps}, systems for analyzing cycling and running routes~\cite{stravaStravaMetro}, and large language models~\cite{touvron2023llama}. 

The data being collected is often personal and sensitive. It contains information about us and our interactions with technologies and people. For example, location data contains information about people's movement across space and time while census data contains information on households, including income. 

Query-based systems (QBSs) are one of the key approaches to safely sharing data. A QBS is an interactive interface that a) allows the data provider to maintain control over a dataset and b) allows an analyst to retrieve answers to queries about a dataset without directly accessing the individual records. For example, a QBS protecting census data can allow an analyst to query the number of people with a salary higher than \$50,000 residing in a given county and return an answer, of say 900. QBS implementations range from web application programming interfaces (APIs) to privacy-preserving SQL engines. They are widely used to share data by industry, academia, and government entities. Examples include QBSs for traffic congestion on the roads by Google Maps~\cite{blogGoogleMaps} and Uber Movement~\cite{uberUberNewsroom}, cycling and running routes by Strava~\cite{stravaStravaMetro} and audience segment attributes by Meta~\cite{facebookIntoFacebook}. Academia~\cite{francis2017diffix, oehmichen2019opal} has proposed sharing data via QBSs with projects such as Airavat~\cite{roy2010airavat} for large-scale parallel computations on sensitive data. Government entities, for example, the Australian Bureau of Statistics (TableBuilder)~\cite{o2008table} and UK's National Health Service (openSAFELY)~\cite{opensafelyOpenSAFELYHome} have used QBSs for sharing census and health data, respectively.

QBSs answer queries by releasing aggregates about the protected dataset.
Releasing aggregates has long been known not to be inherently privacy-preserving~\cite{dwork2015robust,homer2008resolving,dinur2003revealing}. The non-privacy-preserving property is exacerbated by the flexibility given to analysts in QBSs to choose the aggregates themselves. An attacker can send queries highly specific to a target individual, for example, so-called difference queries~\cite{fellegi1972question,denning1979tracker,gadotti2019signal}, which are unlikely to be chosen by the data curator in case of a non-interactive one-time release of aggregates.

To protect privacy, QBSs implement defenses that provide formal privacy guarantees and ad-hoc defenses that do not. Differential privacy (DP)~\cite{dwork2006calibrating} stands as the gold standard for formal privacy guarantees. It aims to protect individual privacy by limiting the impact, measured by a parameter $\epsilon$, called privacy budget, of the inclusion or exclusion of any user's data. The implementation of DP defenses can be challenging in practice. For example, Google Maps~\cite{houssiau2022difficulty} has used weaker, event-level instead of user-level, guarantees, while Amazon's data clean room~\cite{amazonDataCollaboration} and LinkedIn's Audience Engagements~\cite{rogers2020linkedin} regularly (i.e., monthly) reset their budget to accommodate regular data releases, which can invalidate the guarantees in the long run. Even defenses that provide formal privacy guarantees might be at risk, from incorrect implementations~\cite{stadler2022synthetic,tramer2022debugging} to side-channel attacks~\cite{boenisch2021side}. Defenses that do not provide formal privacy guarantees instead rely on adversarial attacks to demonstrate their effectiveness~\cite{francis2017diffix}. This highlights the need to test the privacy guarantees of both types of defenses by using attacks.

There is a trend towards automating the privacy attacks~\cite{shokri2017membership, pyrgelis2017knock,carlini2022membership,cretu2022querysnout} as a response to the increasing flexibility given to the analysts to choose the aggregates from a wider range of options. A privacy attack consists of (1) queries and (2) a rule that combines their answers to infer private information. The level of automation varies from semi- to fully-automated methods. Fully-automated methods automate both the search for queries and the rule~\cite{cretu2022querysnout}, while semi-automated methods require manual efforts in either or both components. All semi-automated methods in the literature manually reduce the possible queries through manual analysis and automate the search for the concrete queries~\cite{gadotti2019signal} or the rule that combines their answers~\cite{pyrgelis2017knock}. The search for queries, without manually reducing the possible queries, is a difficult task, in particular for QBSs that support a wide range of queries.

QuerySnout~\cite{cretu2022querysnout} is, to the best of our knowledge, the only fully-automated method for discovering attacks against a QBS. QuerySnout's automatically discovered attacks consistently outperform, or perform on par with the best-performing known attacks. However, to explore the search space of (multisets of) queries, QuerySnout relies on a computationally expensive evolutionary search technique. Since it maintains a population of multisets of queries in every iteration, its time complexity is proportional to the product of (1) the number of iterations, (2) the number of multisets of queries in the population, and (3) the number of target users. Its ability to find vulnerabilities in QBSs is thus currently limited to a fraction of the attack surface, i.e., query syntax, offered by QBSs and to a highly limited number of target users. These limitations can lead to missed vulnerabilities, such as attacks that rely on expressive query syntax or attacks that only materialize for specific vulnerable users.

\textbf{Contributions.} In this paper, we present QueryCheetah, a method for fast automated discovery of attribute inference attacks against QBSs. At a high level, QueryCheetah moves away from a population of query multisets~\cite{cretu2022querysnout} to a single query multiset. It uses fast locally-informed iterations to search the space of (multisets of) queries. This makes each iteration 450 times faster than previous work while only requiring 25 times more steps, resulting in a speed-up of 18 times.  

In line with previous work, we instantiate QueryCheetah on discovering attribute inference attacks (AIAs) against a real-world QBS, Diffix~\cite{francis2017diffix}, that provides an SQL interface to analysts. The goal of an AIA is to infer a value for a sensitive attribute of the given user of interest.  We formalize the AIAs as a distinguishability privacy game. 

We first show that QueryCheetah outperforms both semi- and fully-automated methods~\cite{gadotti2019signal,cretu2022querysnout}, while being an order of magnitude faster than the state-of-the-art methods. We second show how a) QueryCheetah can discover attacks specific to vulnerable users by attacking many users in a reasonable time, b) how it automatically discovers vulnerabilities in previously unexploited query syntax, and c) how it finds workarounds around defenses~\cite{francis2021specification} developed and deployed to thwart discovered attacks~\cite{cohen2018linear,benthamsgazeLocationTime,differentialprivacyReconstructionAttacks,gadotti2019signal}.

\section{Background}
\subsection{Query-based system}
Let $\mathcal{U}$ denote a universe of users, e.g., users of a service or a country's population. Given a set of users $U \subset \mathcal{U}$, a dataset $D \sim \mathcal{D}$, from a distribution $\mathcal{D}$, consists of the records of these users over a set of attributes $A = \{a_1, ..., a_n\}$, where each attribute $a_i$ can take values in a set $\mathcal{V}_i$. We denote by $s_D=|U|=|D|$ the size of the dataset $D$. For a given user $u \in U$, we denote its record in $D$ by $r_u = (r_u^1, r_u^2, .., r_u^n)$, with $r_u^i$ the user's value for attribute $a_i$. We denote the other possible values for an attribute $a_i$ by $\mathcal{V}_i^u = \mathcal{V}_i \setminus \{r_u^i\}$. Given a subset of attributes $A' =\{a_{i_1}, \ldots, a_{i_k}\} \subset A$ (with $\{i_1, \ldots, i_k\} \subset \{1,\ldots, n\}$ a subset of attribute indexes), we define the projection of a record $r_u$ over attributes $A'$ as $\pi_{A'}: U \rightarrow \mathcal{V}_{i_1}\times\ldots\times \mathcal{V}_{i_k}, \pi_{A'}(u)=(r_u^{i_1}, \ldots, r_u^{i_k})$, which we write more concisely as $r_u^{A'}:=\pi_{A'}(u)$.
\begin{figure}
\centering
\includegraphics[width=\linewidth]{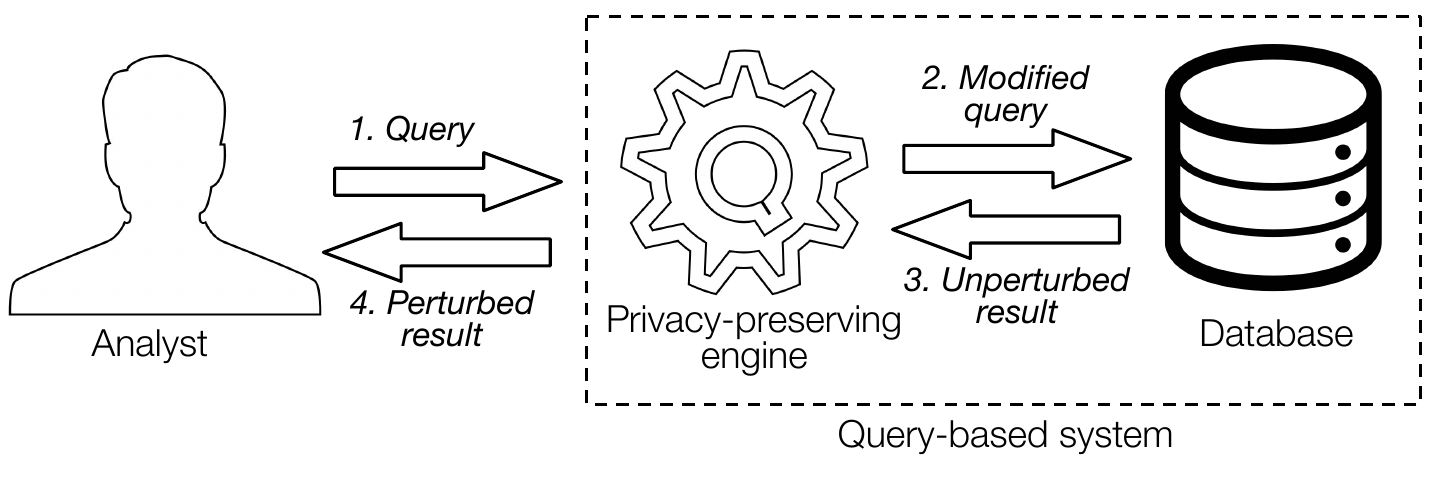}
\caption{Overview of a typical query-based system. An analyst 1) sends a query to the system, which 2) might be modified internally to a form compatible with the database; then 3) the database returns an unperturbed answer, which the 4) system perturbs and returns to the analyst.}
\label{fig:qbs}
\end{figure}

Consider a data curator who wants to enable useful data analyses on a dataset $D$ without releasing individual records. The data curator provides access to the dataset $D$ via a query-based system (QBS), which implements an SQL interface. Analysts submit queries to this interface and retrieve their answers.

We denote by $\mathcal{Q}$ the query syntax supported by the QBS. To answer a query $q\in\mathcal{Q}$, the QBS performs the following steps: the QBS computes first its true answer on the dataset $T(D, q)$\footnote{In practice, there might be an additional query preprocessing step that modifies the query $q$ to $q'$, when, for example, the SQL interface of the QBS is not the same as the SQL interface of the database used to compute the true answer $T(D, q)$. For example, the interfaces might use different SQL dialects. For simplicity, we abstract this step and write $T(D,q)$ and $R(D,q)$ instead of $T(D, q')$ and $R(D, q')$, respectively.}, then (optionally) perturbs it to obtain $R(D, q)$, and finally returns $R(D, q)$ as a final answer. Figure~\ref{fig:qbs} illustrates these steps. We denote by $Y(D,q) \subseteq U$ the userset of query $q$, i.e., the set of users whose records satisfy all conditions in the query $q$. If a QBS receives an unsupported query $q \notin \mathcal{Q}$, in line with the literature~\cite{cretu2022querysnout}, we assume that it returns $0$ as an answer without evaluating it. Returning a $0$ indeed reveals less information than a dedicated response as it introduces uncertainty as to whether the query answer is $0$ or the query is unsupported.

Queries that QBSs with an SQL interface typically support can be denoted as:
\begin{gather}\label{eq:general_syntax}
\begin{aligned}
    \text{SELECT} & \; \text{agg} \;\text{FROM} \; D \\
    \text{WHERE} & \; a_1 \; c_1 \; v_1 \; O_1 \; \ldots \;   O_{n-2} \; a_{n-1} \; c_{n-1} \; v_{n-1} \\ \; & O_{n-1} \; a_n \; c_n \; v_n, 
\end{aligned}
\end{gather}
where: 
\begin{itemize}
    \item agg is an aggregation function $\text{agg} \in \text{AGG}$, such as $\text{AGG} = \{count(), min(a_j), max(a_j), sum(a_j)\}$ for some $j \in \{1,\dots, n\}$, respectively calculating the count of the records, the minimum, maximum or the sum of the values of an attribute $a_j$ of the records satisfying all conditions in the WHERE clause;
    \item $c_i$ is a comparison operator $c_i \in \mathcal{C}$, $i \in \{1, \dots n\}$, such as $\mathcal{C}= \{=, \neq, \perp\}$ where $\perp$ denotes that an attribute is skipped, i.e., the attribute does not appear in the WHERE clause of the query;
    \item $v_i$ is a value, $v_i \in \mathbb{R}$, $i \in \{1, \dots n\}$, and 
    \item $O_i$ is a logical operator, $O_i \in \mathcal{O}$, such as $\mathcal{O} = \{AND, OR\}$, $i \in \{1, \dots n-1\}$.
\end{itemize}  
As an example, when $AGG=\{count()\}$, $\mathcal{C}=\{=, \perp\}$, and $\mathcal{O}=\{AND\}$, the only queries supported by the QBS are contingency tables. 

Providing access to the dataset $D$ through a QBS, instead of releasing individual records, does not suffice to protect the privacy of the users $U$. It has been shown that returning exact answers, or inexact but overly precise answers to too many queries can lead to catastrophic privacy loss in the form of database reconstruction~\cite{dinur2003revealing}. To provide good utility, it is desirable, however, for the perturbed answer $R(D, q), q\in\mathcal{Q}$ to not be too different from the true answer $T(D, q), q\in\mathcal{Q}$. To defend against potential privacy loss, while providing good utility, real-world QBSs implement a range of formal and ad-hoc privacy-preserving mechanisms such as limiting the supported query syntax $\mathcal{Q}$, adding unbiased noise to queries, either bounded~\cite{o2008table} or unbounded~\cite{dwork2006calibrating,francis2017diffix}, seeding of the noise~\cite{o2008table,francis2017diffix}, and query set size restriction meaning that queries selecting too few users, $|Y(D, q)| < T$ for some threshold $T$, are answered with a dummy value such as 0.

\subsection{Diffix}\label{background:diffix}
In this paper, we focus on a real-world QBS, Diffix~\cite{francis2017diffix}, which implements a complex combination of privacy-preserving mechanisms, whose privacy loss can be difficult to manually test in practice. It is the most heavily studied and developed QBS that does not provide formal privacy guarantees. 

Diffix has used discovered attacks to patch and improve the system over time. The authors have organized two bounty programs~\cite{aircloakAttackChallenge, aircloakWorldsOnly}, where experts were invited to adversarially test the privacy guarantees of the system and get monetary prizes in return. Four manual or semi-automated vulnerabilities were discovered~\cite{aircloakSecurityAircloak}: a membership inference attack (MIA)~\cite{benthamsgazeLocationTime} based on earlier work \cite{pyrgelis2017knock}, two reconstruction attacks~\cite{cohen2018linear, differentialprivacyReconstructionAttacks} based on earlier work as well~\cite{dinur2003revealing}, and an attribute inference attack (AIA)~\cite{gadotti2019signal}. A fully-automated method has discovered AIAs against Diffix with stronger inference capabilities~\cite{cretu2022querysnout}. When patching the system against the discovered vulnerabilities, additional defenses were introduced, against which new vulnerabilities were discovered, leading to multiple versions of the system over time~\cite{francis2018diffix, francis2020specification, francis2021specification}. In this paper, we focus on up to the last version of Diffix against which, to the best of our knowledge, no known attacks exist, Diffix-Dogwood.\footnote{In the latest version of Diffix, Diffix-Elm \cite{francis2022diffix}, only a highly limited syntax is supported. For example, conditional queries are forbidden, i.e., queries with a \textit{WHERE} clause, on which all known AIAs against it rely on. This limitation drastically reduces the general-purpose utility of the system. }

We refer to the defenses that were introduced in Diffix's first version as \textit{main defenses}, while the defenses that were introduced in later versions to thwart discovered attacks as \textit{mitigations}.

\subsubsection{Diffix's main defenses}
Diffix's main defenses are: query-set size restriction, unbiased unbounded noise, and answer rounding. 
\begin{enumerate}
    \item \textit{Query-set size restriction}: Diffix restricts releasing the answer of a query $q$ if its true answer $T(D, q)$ is lower than $2$ or a noisy threshold $T$, $T(D, q) < max(2, T)$, distributed as $T \sim N(4, 0.5)$, where $N(\mu, \sigma)$ denotes the Gaussian distribution with mean $\mu$ and standard deviation $\sigma$. 
    \item \textit{Unbiased unbounded noise}: Non-restricted queries are perturbed by adding two layers of noise, static and dynamic, to the true answer $T(D,q)$ for each filtering condition $a_i\;c_i\;v_i$, $i \in \{1, \dots, n\}$ in the \textit{WHERE} clause of the query. The static $N_s^i$ and the dynamic $N_d^i$ noise terms are distributed as $N_s^i, N_d^i \sim N(0, 1), i\in\{1,\dots,n\}$.
    \item \textit{Answer rounding}: Finally, the noisy answer of non-restricted queries is rounded to the nearest integer, if the type of the true answer is an integer. Otherwise, the answer is not rounded.
\end{enumerate}

Diffix uses a seeded pseudo-random number generation (PRNG) for sampling the threshold $T$ and the static and dynamic noise terms $N_s^i, N_d^i, i\in\{1,\dots,n\}$.  The seeding ensures that if a query $q$ is received more than once, Diffix will respond with the same answer every time. For seeding the PRNG, Diffix is initialized with a secret salt $p$ and an attribute $a_0$ that uniquely identifies all user records in $D$, $\forall u,v \in U, r_u^0 \neq r_v^0, u \neq v$. Diffix seeds the PRNG for generating $T$ with the secret salt $p$ and the userset $Y(D, q)$. The sampling of the static noise, $N_s^i$, corresponding to the condition on the attribute $a_i$, is seeded with the secret salt $p$ and the syntax of the condition itself, i.e., with the attribute $a_i$, the comparison operator $c_i$, and the value used $v_i$. The seed for the dynamic noise $N_d^i$ additionally includes the userset $Y(D, q)$ by using a bit-wise XOR over the attribute $a_0$ of all users in the userset $XOR(r_{u_1}^0, \dots, r_{u_{|Y(D, q)|}}^0), Y(D, q) = \{u_1, \dots, u_{|Y(D, q)|}\}$. 

Overall, the perturbed answer $R(D, q)$ has the following form:
\[ R(D, q) = \begin{cases} 
      0 & T(D,q)\leq max(2, T) \\
      round(T(D,q) + \sum\limits_{i=1}^n N_s^i + \sum\limits_{i=1}^n N_d^i) & otherwise
   \end{cases}
\]

\subsubsection{Query syntax}\label{sec:diffx_query_syntax}
The privacy guarantees of the three mechanisms have only been tested within a limited query syntax, $\mathcal{Q}_{lim}$, by previous work developing attribute inference attacks (AIAs). The limited syntax $\mathcal{Q}_{lim}$ allows queries where:
\begin{itemize}
    \item $\text{AGG} = \{count()\}$;
    \item $\mathcal{C}= \{=, \neq, \perp\}$;
    \item $v_i = r_u^i$, $i \in \{1, \dots n-1\}$ and $v_n \in\{0,1\}$; 
    \item $\mathcal{O} = \{AND\}$.
\end{itemize}  

Diffix, however, supports a query syntax richer than $\mathcal{Q}_{lim}$~\cite{francis2020specification,francis2021specification}. This leaves a large part of the syntax currently unexplored whether or not it leads to privacy vulnerabilities. 

The limited syntax $\mathcal{Q}_{lim}$ can be extended to a richer syntax of supported counting queries,
$\mathcal{Q}_{ext}$. The syntax $\mathcal{Q}_{ext}$ extends $\mathcal{Q}_{lim}$ along $4$ axes, $D_1, D_2, D_3$, and $D_4$, $\mathcal{Q}_{ext} = \mathcal{Q}_{lim} \cup \{D_1, \dots, D_4\}$:
\begin{enumerate}
    \item[($D_1$)] \textit{Allowing any value}: Allow conditions to compare to any value $v_i$ that is not necessarily the target user's value $r_u^i$. When axis $D_1$ is used with any other axes $D_2, D_3$ or $D_4$, $v_i$ can be a pair of values. Formally, $D_1$ extends the syntax to allow comparisons to any real value or pair of real values, $D_1 := v_i \in \mathbb{R} \cup (\mathbb{R}\times\mathbb{R})$, where $\mathbb{R}$ is the set of real numbers, $i \in \{1, \dots, n\}$.
    \item[($D_2$)] \textit{Allowing BETWEEN}: Allow the comparison operator BETWEEN, $\textit{BETWEEN} \in \mathcal{C}$. This allows conditions $a_i$ $BETWEEN$ $(v_{i,1}, v_{i,2}),$ $v_{i,1} < v_{i,2}$ that compare whether the value of the attribute $a_i$ is in the interval $(v_{i,1}, v_{i,2})$. Diffix allows intervals $(v_{i,1}, v_{i,2}), v_{i,1} < v_{i,2}$, with a width, $w = v_{i,2} - v_{i,1}$, that falls in the infinite set $w \in \{\dots, 0.1, 0.2, 0.5,$ $1, 2, 5,$ $10, 20, 50, \dots\}$ and offset $v_{i,1}$ that falls on an even multiple of the width, $v_{i,1} = 2kw$, or an even multiple plus $\frac{1}{2}$ of the width, $v_{i,1}=(2k + \frac{1}{2})w$, for some integer $k \in \mathbb{Z}$. 
    \item[($D_3$)] \textit{Allowing IN}: Allow the comparison operator \textit{IN}, $\textit{IN} \in \mathcal{C}$. This allows conditions $a_i\;IN\;\{v_{i,1}, v_{i,2}\}$ that compare whether the value of the attribute $a_i$ is in the set $\{v_{i,1}, v_{i,2}\}$. While Diffix allows for sets with more elements, here we focus only on two-element sets, in particular, $v_{i,1} = r_u^{a_i}$ and $v_{i,2} \in \mathcal{V}_i^u$. 
    \item[($D_4$)] \textit{Allowing NOT IN}: Allow the comparison operator \textit{NOT IN} operator, $\textit{NOT\;IN} \in \mathcal{C}$. This allows conditions $a_i$ $NOT\;IN$ $\{v_{i,1}, v_{i,2}\}$ that compare whether the value of the attribute $a_i$ is in \textit{not} the set $\{v_{i,1}, v_{i,2}\}$. We use the same domains for $v_{i,1}$ and $v_{i,2}$ as in $D_3$.
\end{enumerate}

The extended syntax $\mathcal{Q}_{ext}$ is a much broader space than $\mathcal{Q}_{lim}$. Table~\ref{tab:sizes} presents the numbers of possible multisets of queries in $\mathcal{Q}_{ext}$ and $\mathcal{Q}_{lim}$, and thus the number of possible attacks in those syntaxes, and compares them with the space sizes of other problems.

\begin{table}[t]
    \centering
    \begin{tabular}{lr}
        Problem & Count\\ \hline \hline
        Atoms in the observable universe \cite{ade2016planck} & $\approx 10^{80\hphantom{1}}$\\ \hline
        Possible chess games \cite{shannon1950xxii} &  $\approx 10^{120}$\\ \hline
        Possible multisets of $100$ queries in $\mathcal{Q}_{lim}$  & \multirow{2}{*}{$1.33 \cdot 10^{131}$} \\
        (with $\mathcal{V}_n=\{1\}$)~\cite{cretu2022querysnout} & \\ \hline
        Possible multisets of $100$ queries in $\mathcal{Q}_{ext}$\tablefootnote{We keep the same assumptions as QuerySnout's estimation of $m=100$ queries in a multiset and $n=6$ total attributes. Assuming we have 5 ordinal attributes that can be used with any comparison operator, including \textit{BETWEEN}, and conservatively assuming that there are $15$ possible ranges we can ask for \textit{BETWEEN} and $15$ sets for \textit{IN} and \textit{NOT IN} operators, there are $48^5\cdot5$ possible queries. This makes up for a search space of size $\binom{48^5\cdot5 + m - 1}{m} \approx 3.53 \cdot 10^{752}$.}  & \multirow{1}{*}{$3.53 \cdot 10^{752}$}  
    \end{tabular}
    \caption{Search space sizes.}
    \label{tab:sizes}
\end{table}

\subsubsection{Diffix's mitigations}~\label{sec:mitigations_diffix}
To thwart discovered attacks, Diffix implements additional defenses, which we refer to as mitigations. We have identified four mitigations for the counting queries in the extended syntax $\mathcal{Q}_{ext}$:
\begin{enumerate}
    \item \textit{Isolating attributes}: Forbid queries of at least one condition $a_i\;c_i\;v_i$ with a comparison operator $\neq$ or \textit{IN}, $c_i \in \{\neq, IN\}$ $i\in\{1,\dots,n\}$ where 80\% or more of values for the attribute $a_i$ uniquely identify the users. Such an attribute $a_i$ is called an isolating attribute.  
    \item \textit{Shadow table}: Forbid queries of at least one condition with a comparison operator $\neq$ or \textit{IN}, $c_i \in \{\neq, IN\}$ $i\in\{1,\dots,n\}$ for values $v_i$ that are not among the top $200$ most frequent values for attribute $a_i$ that appear at least for $10$ distinct users. Diffix creates at initialization a so-called shadow table that stores these frequent values for each attribute that can be used with comparison operators $\neq$ and \textit{IN}.
    \item \textit{Noise when no conditions}: Add noise to the query with no filtering conditions in the \textit{WHERE} clause, $c_i = \perp, \forall i\in\{1,\dots,n\}$. This ensures that this query is also perturbed, as the main noise addition mechanism adds noise per condition.
    \item \textit{Dynamic noise seed}: Seed the dynamic noise terms $N_d^i, i\in\{1,\dots ,n\}$, which depends on the userset $Y(D,q)$, slighly differently. Namely, seed the PRNG by using statistics (minimum, maximum, and count) of the values of the attribute $a_0$, $min(X), max(X), |X|, X=\{r_u^0 | u \in Y(D,q)\}$, instead of aggregating them by using bit-wise XOR. 
\end{enumerate}

\subsection{Threat model for attribute inference attacks}
In this paper, we evaluate the privacy guarantees of Diffix in the context of attribute inference attacks (AIAs). We follow the threat model that has been used by existing AIAs against Diffix~\cite{gadotti2019signal,cretu2022querysnout}.

AIAs aim to infer the value of a sensitive attribute for a target user $u$. Without loss of generality, we assume that the last attribute, $a_n$, is sensitive, and following the literature, we assume for simplicity that it is binary. The AIA thus aims to infer the binary value $r_u^n$. 

Although we focus on AIAs, note that our method can be extended to membership inference attacks (MIAs). MIAs aim to infer the presence of a given user record $r_u$ in the dataset, i.e., whether $r_u \in D$, for some $u \in \mathcal{U}$.

\textbf{Access to the query-based system}.
The threat model considers an attacker who has access a) to an instantiation of Diffix protecting a dataset $D$ and b) to Diffix's software.

The attacker can access the protected dataset $D$ only through sending queries to the Diffix instantiation. We refer to this Diffix instantiation as the target QBS. Although Diffix allows an unlimited number of queries, we assume that the attacker can send at most $m$ queries to the target QBS. Since Diffix logs all received queries~\cite{boenisch2021side}, using a low number of queries, typically tens or low hundreds, can help to avoid detection.

We assume that the attacker can use Diffix's software as a black-box executable, in line with the literature of fully-automated attacks~\cite{cretu2022querysnout}. Semi-automated attacks assume a white-box access to the software, which allows to manually study the system in detail, such as the noise addition mechanism~\cite{gadotti2019signal}. Note that the white-box access here refers to access only to the software, and not to any values specific to the target QBS, such as the secret salt $p$.

\textbf{Access to auxiliary knowledge}.
The attacker has auxiliary knowledge a) about the target user $u$ and b) about the dataset $D$.

The attacker has access to a projection of the target record $r_u$ on a subset of attributes $A' \subseteq A \setminus \{a_n\}$, i.e., the attacker knows $r_u^{A'}$, and they also know that the target user is uniquely identifiable on $A'$, $\forall v \in U, v\neq u, r_v^{A'} \neq r_u^{A'}$. 

We denote by $K$ the attacker's auxiliary knowledge about the dataset $D$. We follow the literature of fully-automated attacks~\cite{cretu2022querysnout} and instantiate $K$ as knowledge of an auxiliary dataset, $K= D_{aux}\sim\mathcal{D}_{aux}$ of a distribution $\mathcal{D}_{aux}$, $D_{aux}\sim \mathcal{D}_{aux}$, similar to the distribution $\mathcal{D}$ of the protected dataset $D$. Semi-automated attacks instantiate $K$ as knowledge of the subsets of attributes for which the target record is unique~\cite{gadotti2019signal}.

\textbf{Attacker's goal}
The attacker's goal is to discover a multiset $S$ of $m$ queries, $S=\{q_1, \dots, q_m\}$ and a rule $V$ to combine their answers by the target QBS, $V(R(D, q_1), \dots, R(D, q_m))$, to a prediction of the sensitive value $r_u^n$. 

\subsection{Semi-automated AIA against Diffix: Differential noise-exploitation attack}\label{sec:gadotti}
Gadotti et al.~\cite{gadotti2019signal} introduced a semi-automated AIA against Diffix, called a differential noise-exploitation attack. It exploits Diffix's layered noise addition mechanism. The attack uses manually-identified pairs of queries, $(q_1, q_2), q_1,q_2 \in\mathcal{Q}_{lim}$, whose usersets are either identical or differ in one record, depending on the value of the target user's sensitive attribute. In particular, it uses:
\begin{gather}\label{eq:difference_queries}
\begin{aligned}
    q_1 := \text{SELECT} & \; \text{count()} \;\text{FROM} \; D \\
    \text{WHERE} & \; a_{i_1} \; \neq \; r_u^{i_1} \; AND \; a_{i_2} \; = \; r_u^{i_2} \; AND \; \ldots  \\
    & \ldots \;   AND \; a_{i_{l}} \; = \; r_u^{i_{l}} \; AND \; a_n \; = \; v_n, \\
    q_2 := \text{SELECT} & \; \text{count()} \;\text{FROM} \; D \\
    \text{WHERE} & \; \hphantom{a_{i_1} \; \neq \; r_u^{i_1} \; AND \;} a_{i_2} \; = \; r_u^{i_2} \; AND \; \ldots  \\
    & \ldots \;   AND \; a_{i_{l}} \; = \; r_u^{i_{l}} \; AND \; a_n \; = \; v_n, 
\end{aligned}
\end{gather}
for $v_n\in\{0,1\}$ and a subset of attributes $A''=\{i_1, \dots, i_l\}$ of the attributes known to the attacker $A'$, $A'' \subseteq A'$, on which (1) the target user $u$ is uniquely identifiable,  $\forall v \in U, v\neq u, r_v^{A''} \neq r_u^{A''}$, and (2) both queries $q_1$ and $q_2$ are not bucket suppressed. This attack uses on the fact that the target user $u$ is the only user in the userset of $q_2, u\in Y(D, q_2)$ who is not in the userset of $q_1, u \notin Y(D, q_1)$, $\forall v\in U, v\neq u, v\notin Y(D, q_2) \lor v \in Y(D, q_1)$. By calculating the difference in the query answers $\Delta = R(D, q_2) - R(D, q_1)$ most of the static noise terms cancel out and the dynamic noise terms depend on the usersets of the queries that condition on the sensitive value. The difference $\Delta$ is distributed as a Gaussian $\Delta\sim N(0,2)$ if $r_u^n = 1-v_n$ and as a $\Delta\sim N(1, 2l+2)$ if $r_u^n = v_n$.\footnote{This analysis assumes an idealistic scenario that Diffix does not round the query answers. However, the variances of the two cases still differ even after rounding.} To distinguish between the distributions of these two cases, the attack employs a likelihood ratio test. 

The authors perform an automated search over the space of attribute subsets $A''$ to find the subsets that fulfill conditions (1) and (2). Note that, we here refer to the overall attack pipeline as only semi-automated because there are two parts discovered through manual analysis: the core vulnerability (queries) is identified manually by the authors, allowing them to restrict the search space, and the rule to combine the queries is manually crafted (likelihood ratio attack).

Note that Gadotti et al.~\cite{gadotti2019signal} extended the differential attack by appending filtering conditions $a\;c\;v$, $a,c,v\in A \times \mathcal{O} \times \mathbb{R}$ to queries $q_1$ and $q_2$, that do not change their usersets (e.g., appending the condition $years\_at\_company\neq 10$ to a query containing $employment\_year=2024$). Following the literature on automated attacks~\cite{cretu2022querysnout}, we do not include the extension in our comparison as crafting the conditions relies on domain knowledge.

\subsection{Fully-automated AIA against Diffix: QuerySnout}\label{sec:querysnout}

Cretu et al.~\cite{cretu2022querysnout} proposed the only fully-automated method for discovering AIAs against Diffix, called QuerySnout. QuerySnout automates both parts that were manual in previous work: (1) the search for a rule to combine query answers in predicting the sensitive value and (2) the search for candidate multisets of queries.\footnote{QuerySnout's code is publicly available at \url{https://github.com/computationalprivacy/querysnout} and we reuse it when comparing to state-of-the-art approaches.} By devising the rule that combines the answers, QuerySnout estimates the likelihood that a given multiset of queries constitutes an attack, called the fitness of the multiset.

\textbf{Fitness: combining the query answers}. QuerySnout extends to AIAs an existing technique for estimating the vulnerability to MIAs by Pyrgelis et al.~\cite{pyrgelis2017knock}.

\begin{figure*}[t]
    \centering
    \includegraphics[width=0.72\linewidth]{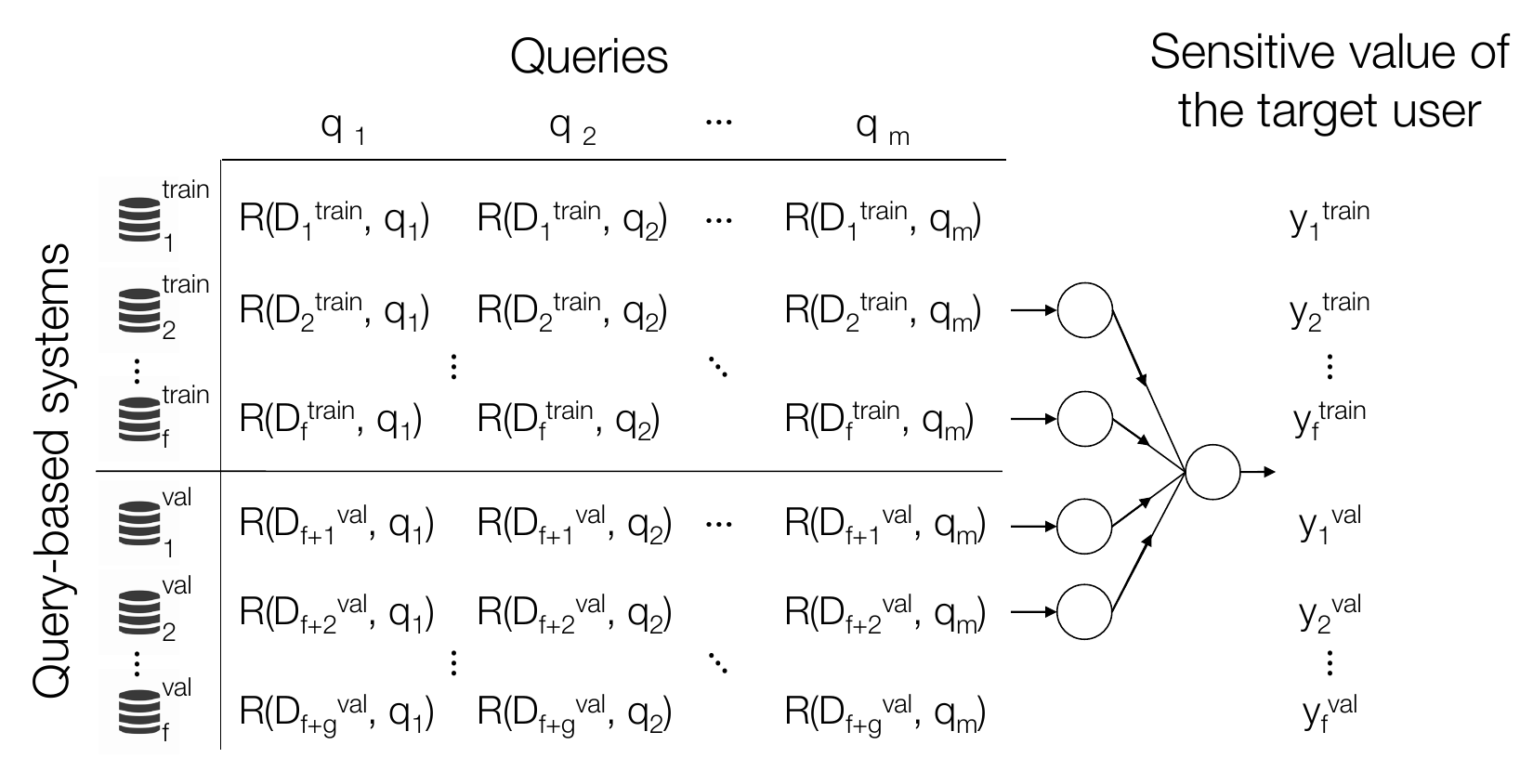}
    \caption{Overview of the pipeline used by the attacker in QuerySnout~\cite{cretu2022querysnout} to estimate the fitness of a multiset of queries.}
    \label{fig:setup}
\end{figure*}

The attacker uses the auxiliary knowledge $K=D_{aux}$ and performs the following $9$ steps to estimate the fitness of a query multiset $S=\{q_1,$ $\dots,$ $q_m\}$:
\begin{enumerate}
    \item Split the auxiliary dataset $D_{aux}$ in two equal partitions $D_{aux}^{train}$ and $D_{aux}^{val}$. 
    \item Uniformly at random sample without replacement $z$ records from $D_{aux}^{train}$, $\{r_{v_1}, \dots r_{v_z}\}$, project them on $A'$,  $\{r_{v_1}^{A'}, \dots r_{v_z}^{A'}\}$ and add the target user's record $r_u^{A'}$ to create a collection of $z+1$ records $T=\{r_{v_1}^{A'}, \dots r_{v_b}^{A'}, r_u^{A'}\}$. The goal of this step is to simulate the QBS behavior on similar datasets having different values of the sensitive attribute.
    \item For each of the $z+1$ records in $T$, independently sample a Bernoulli distribution, $Bernoulli(0.5)$, and create a dataset $D_1^{train}=\{r_{v_1}^{A'} \cup \{b_1\}, \dots r_{v_z}^{A'} \cup \{b_z\}, r_{u}^{A'} \cup \{b_u\}\}$ over attributes $A' \cup \{a_n\}$, where $b_1, \dots b_z, b_u \sim Bernoulli(0.5)$. Denote $b_u$ by $y_1^{train}=b_u$.
    \item Repeat steps (2) and (3) to create $f$ datasets $D_1^{train}, \dots, D_f^{train}$ and a vector of $f$ values for the target user $(y_1^{train}, \dots, y_f^{train})$.
    \item Repeat steps (2) and (3) $g$ times for $D_{aux}^{val}$ to create $D_1^{val}, \dots, D_g^{val}$ and $(y_1^{val}, \dots, y_g^{val})$.
    \item Protect each of $f+g$ datasets with the QBS by using the executable software they have access to. Each of the $f+g$ instantiations of the QBS uses different values for the secret salt $p$.
    \item Evaluate the $m$ queries in $S$ to each of the $f+g$ QBSs and obtain an $(f+g)$ vectors of  $m$ query responses: $[R(D_1^{train}, q_1),$ $\dots,  R(D_1^{train}, q_m)], \dots, [R(D_{f+g}^{val}, q_1), \dots,  R(D_{f+g}^{val}, q_m)]$.
    \item Finally, train a logistic regression model to predict the sensitive value $y_i^{train}$ given the vector with $m$ query answers, $[R(D_i^{train}, q_1),$ $\dots,  R(D_i^{train}, q_m)]$, from the QBS protecting $D_i^{train}$,  as shown in Figure~\ref{fig:setup}. The answers on the $g$ QBSs protecting the datasets of step 5. are used for validation purposes.
    \item Compute the multiset fitness $\bar{F} = min(acc_{train}, acc_{val})$, where $acc_{train}$ and $acc_{val}$ denote the accuracy of the ML model predicting the sensitive value on QBSs from $D_{aux}^{train}$ and QBSs from $D_{aux}^{val}$, respectively. 
\end{enumerate}

\textbf{Finding candidate multisets of queries}. QuerySnout proposes an evolutionary search technique with custom mutations to search for a multiset of queries with a high fitness value. It maintains a population $P_i$ of $P$ multisets of $m$ queries,  $P_i=\{S_{1,i}, \dots, S_{P,i}\}$ in each iteration $i$, where $S_{j,i} = \{q_{j,i,1}, \dots, q_{j,i,m}\}, \forall j\in \{1,\dots, P\}$, for $q_{j,i,1}, \dots, q_{j,i,m}\in\mathcal{Q}_{lim}$ where $v_n=1$. 

First, it initializes the population $P_0$ by constructing multisets of randomly sampled queries. Then, it performs an evolutionary search over $I$ iterations. Finally, the attacker selects the multiset in $P_I$ with the highest fitness value and attacks the target QBS by sending the queries in it. Appendix~\ref{sec:addition_details_about_querysnout} describes more details.

\section{Formalizing AIA against a QBS as a privacy game} \label{subsec:game} 
In this section, we formalize a targeted attribute inference attack against a user $u$, given the multiset of $m$ queries $S=(q_1, ..., q_m)$ as a privacy game. The game tests the strength of an attack $F(S)$ given with the query multiset $S$. It is played between two players, Alice in the role of the defender, i.e., data curator, and Bob in the role of the challenger, i.e., attacker. We use the game to estimate the severity (if any) of an AIA that uses the multiset $S$. 

The game is parameterized by: a query multiset $S$, a data distribution $\mathcal{D}$, a dataset size $s_D$, a target user's $u$ partial record $r_u^{A'}$, a sensitive attribute $a_n\in A \setminus A'$, and an adversary's side knowledge $K=D_{\text{aux}}$.

\begin{thm} (QBS\_AIA(S, $\mathcal{D}$, $s_D$, $r_u^{A'}$, $a_n$ $D_{\text{aux}}$, QBS, Alice, Bob))
The game is repeated $R$ times. In each repetition, Alice goes first, and takes three steps:
\begin{enumerate}
    \item She samples a dataset of $s_D-1$ records from $\mathcal{D}$, $\{r_{u_1}, \dots, r_{u_{s-1}}\}$, projects them on $A'$, $\{r_{u_1}^{A'}, \dots, r_{u_{s-1}}^{A'}\}$ and adds the target user record $r_u^{A'}$, $D_i = \{r_{u_1}^{A'}, \dots, r_{u_{s-1}}^{A'}, r_u^{A'}\}$ for the $i^{th}$ repetition of the game.
    \item For each of the $s_D$ records in $D_i$, she samples $b_j\sim Bernoulli(0.5),$ $j\in\{1,\dots,s\}$ and creates a ``shadow dataset'' $D_i^{test}$, $D_i^{test} = \{r_{u_1}^{A'} \cup \{b_1\}, \dots, r_{u_{s-1}}^{A'} \cup\{b_{s-1}\}, r_u^{A'} \cup \{b_s\}\}$. The values $b_j$ represent values for the sensitive attribute $a_n$. 
    She stores the sensitive value assigned to the target user $u$, $b_s$ as $y_i^{test} = b_s$.
    \item Finally, she protects the dataset $D_i^{test}$  with a QBS.
\end{enumerate}

Once Alice finishes her turn, Bob goes next with two steps:
\begin{enumerate}
    \item Bob asks the queries in $S$ to Alice's QBS and gets query answers back $\{R(D_i^{test}, q_1),\dots, R(D_i^{test}, q_m)\}$.
    \item From the answers, he predicts the sensitive value $y_i^{test}$. In the $i^{th}$ repetition of the game, Bob wins if he correctly predicts the sensitive value $y_i^{test}$, $w_i=1$. Otherwise, he loses, $w_i=0$. 

\end{enumerate}

\end{thm}

In line with previous research \cite{gadotti2019signal,cretu2022querysnout}, with the second Alice's step, the game allows breaking the correlations the sensitive attribute has with the other attributes. This, however, helps in measuring the privacy leakage of the QBS isolated from other confounding variables. This also helps to establish a baseline, which in this case is a random coin flip guess with 50\% accuracy.

After $R$ repetitions, the fraction of Bob's wins are reported. We denote by $F(S) = \frac{\sum_i w_i}{R}$ the accuracy of the multiset $S$. If Bob has won many of the repetitions, the query multiset $S=(q_1, \dots, q_m)$ is likely to constitute a potential privacy attack.

In practice, the distribution $\mathcal{D}$ is instantiated as a dataset of size larger than $s_D$ and the datasets $D_1^{test}, \dots, D_R^{test}$ are created by sampling records from it uniformly without replacement.
The datasets $D_1^{test},\dots, D_R^{test}$ simulate the protected dataset $D$ sampled from $\mathcal{D}$.

\section{Methodology}
\label{sec:methodology}

\begin{figure*}[!t]
    \centering
    \includegraphics[width=0.72\linewidth]{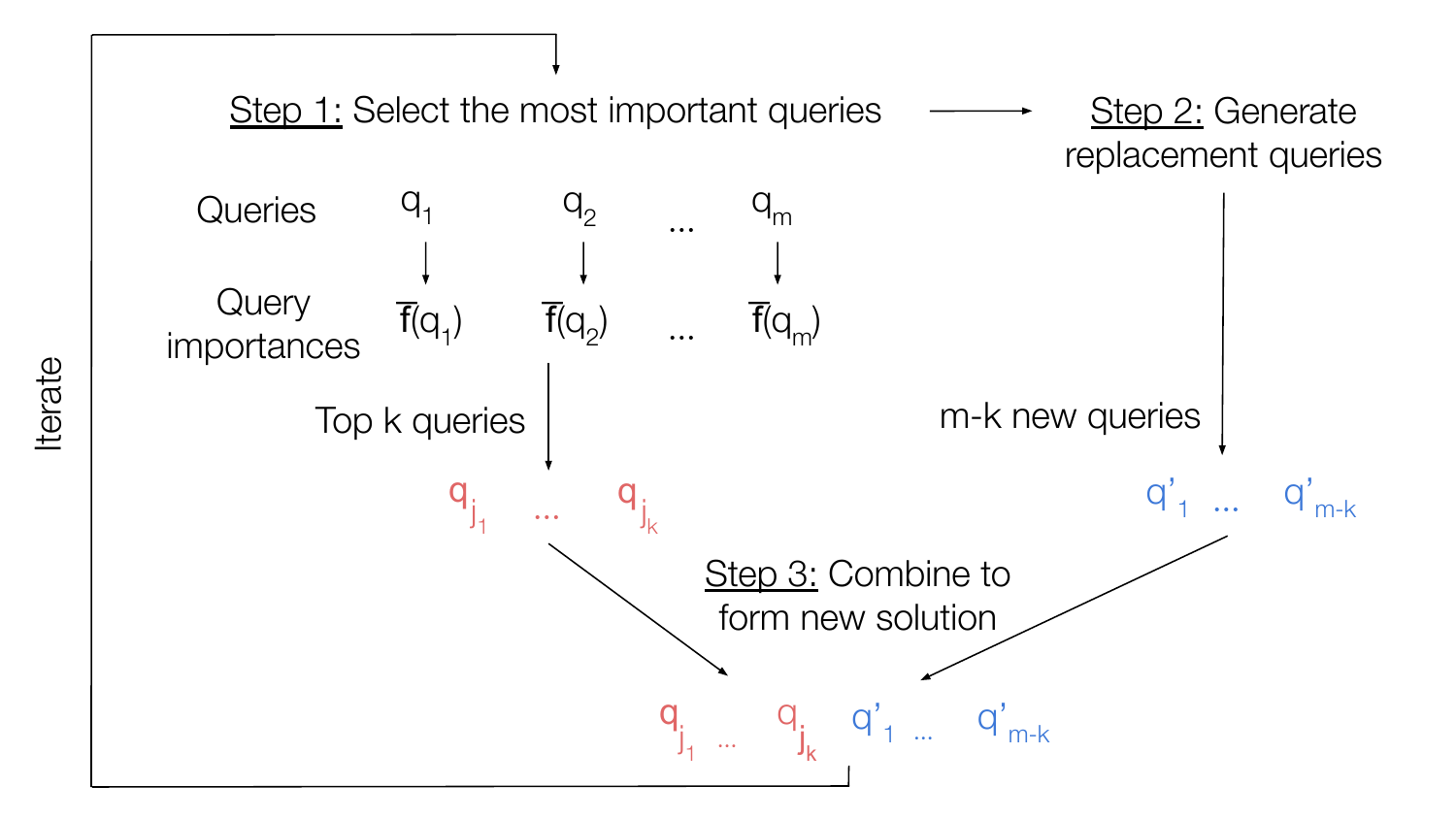}
    \caption{Illustration of QueryCheetah's local search in one stage.}
    \label{fig:method_limited}
\end{figure*}
In this section, we describe QueryCheetah, our novel method for fast automated discovery of AIAs against QBSs. Our goal is to discover a multiset of $m$ queries whose answers by the target QBS leak the sensitive attribute of the target user $u$. QueryCheetah uses a multi-stage local-search technique. In each stage, QueryCheetah performs a local search. Each stage corresponds to a specific query syntax. In the first stage, it starts by searching a multiset of queries in the limited syntax $\mathcal{Q}_{lim}$. In each next stage, it explores a richer syntax. In the final stage, it explores the extended syntax $\mathcal{Q}_{ext}$. Overall, it performs $I$ iterations, divided into $b$ stages of $I_1, \dots, I_b$ iterations each, $I=I_1 + \dots + I_b$.

\textbf{Local search}. QueryCheetah performs a local search within each stage, searching for a multiset of $m$ queries with high fitness $\bar{F}$. In the context of the search, we refer to the multisets of queries as solutions. 

Local search methods~\cite{chen2016multi} select the next solution $S_{i+1}$ as a solution obtained from the previous one $S_i$, $S_{i+1} = M(S_i)$, where $M$ denotes the method that selects the next best iterative step. The goal of the method $M$ is to select $S_{i+1}$ similar to $S_i$ that improves the fitness, $\bar{F}(S_{i+1}) > \bar{F}(S_i)$. QueryCheetah parametrizes $M$ with the query syntax $\mathcal{Q}$ that is explored in the current stage, $S_{i+1} = M(S_i, \mathcal{Q})$. We define $M$ to construct $S_{i+1}$ as a union of (1) retained $k$ queries from $S_i$, $r(S_i) = \{q_{j_1, i}, \dots, q_{j_k, i}\}$ by using a method $r$ to obtain the query indices $J = \{j_1, \dots, j_k\}$ and (2) generated $m-k$ queries $g(S_i, \mathcal{Q}) = \{q_{1,i}', \dots, q_{m-k,i}'\}, q_{j,i}', j\in\{1, \dots, m-k\}$ by using a method $g$ for generating a query in syntax $\mathcal{Q}$. The goal of the query selection $r(S_i)$ and generation $g(S_i, \mathcal{Q})$ methods is to construct $S_{i+1}$ to improve the fitness $\bar{F}(S_{i+1}) \geq \bar{F}(S_i)$ while at the same time being computationally cheap. 

We instantiate $r(S_i)$ to assign importance scores $\bar{f}(q_{j,i})$ to every query in $S_i$ $q_{j,i}\in S_i, j\in\{1,\dots,m\}$ and select the $k$ queries with highest scores, $top\_k(S_{i}, \bar{f}, k)=argmax_{\{j_1, \dots, j_k\}}\sum_{j \in \{j_1, \dots, j_k\}}\bar{f}(q_{j,i})$. As the multiset fitness $\bar{F}(S_i)$ is calculated by training a logistic regression model that uses queries as features (see Section~\ref{sec:querysnout}), we use feature importance scores as query importance. For simplicity and computational efficiency, we opt for a model-specific feature importance score, given by the absolute values of the corresponding coefficient of the logistic regression.

We instantiate $g(S_i, \mathcal{Q})$ as a generator of random queries in query syntax $\mathcal{Q}$. A random query is generated by iterating over the attributes $A$. For each attribute $a_i, i\in\{1,\dots,n\}$ we sample two elements: first a comparison operator $c\in\mathcal{C}$ supported in the syntax $\mathcal{Q}$ and second a random value $v \in \mathbb{R} \cup \mathbb{R}\times\mathbb{R}$ corresponding to the comparison operator $c$. Appendix~\ref{sec:random_query} describes the method in more detail. Note that our instantiation choice for $g(S_i, \mathcal{Q})$ does not depend on the multiset $S_i$. It can however be extended to incorporate such dependence. For example, extending it to copy queries from $S_i$ might be useful when attacking a target QBS which, in contrast with Diffix, does not guarantee returning the same answer if a query is received more than once.

\begin{algorithm}[t]

\SetCommentSty{mycommfont}
\SetAlgoNoLine

\SetKwInput{KwInput}{Input}                
\SetKwInput{KwOutput}{Output}              
\DontPrintSemicolon
  
  \KwInput{Query syntax $\mathcal{Q}$, \;
  \hphantom{Input:  } Starting multiset of queries $S_{start}$. \;
  \hphantom{Input:  } Query multiset size $m$, \;
  \hphantom{Input:  } Number of attributes $n$, \;
  \hphantom{Input:  } Target user's partial record $r_u^{A'}$, \;
  \hphantom{Input:  } Number of iterations $num\_iters$, \;
  \hphantom{Input:  } Number of queries to keep $k$, \;
  \hphantom{Input:  } Solution fitness function $\bar{F}$, \;
  \hphantom{Input:  } Query importance function $\bar{f}$
  }
  \KwOutput{Multiset of queries}
  \eIf{$S_{start} \neq \emptyset$} {
   $S_0$ $\gets$ $S_{start}$ \tcp*{Initialize if explicitly given a starting query multiset}
  }
  {
   $S_0$ $\gets$ [\textit{random\_query}($\mathcal{Q}$, $n$) for $j=1$ to $m$] \tcp*{Sample $m$ random queries from $\mathcal{Q}$}
  }
  
  \For{$i=0$ to num\_iters} {
    $S'_i$ $\gets$ top\_k($S_i$, $\bar{f}$, $k$) \tcp*{select the top-$k$ queries with largest $\bar{f}$ values to keep for next iteration}
    $S''_i$ $\gets$ $S_i$ $\setminus$ $S'_i$ \tcp*{discard the rest $m-k$}
    $S_{i+1}$ $\gets$ $S'_i$ $\cup$ [\textit{random\_query}($\mathcal{Q}$) for $j=1$ to $m-k$] \tcp*{replace the discarded queries}
  }
  $best\_iteration$ $\gets$ $argmax_i(\bar{F}(S_i))$\; 
  \Return $S_{best\_iteration}$
\caption{\textsc{QueryCheetahSyntaxSubsetSearch}}
\label{algo:limited}

\end{algorithm}

Figure~\ref{fig:method_limited} illustrates QueryCheetah's local search in one stage exploring multisets in syntax $\mathcal{Q}$ and Algorithm~\ref{algo:limited} presents the detailed pseudocode. We start from an initial multiset $S_0$ (lines 1-5). Iteratively (line 6), we perform the three steps of $M$, illustrated in Figure~\ref{fig:method_limited}, to obtain a solution $S_i, i>1$: (1) select the $k$ queries from $S_{i-1}$ with the highest query importances by using $r(S_{i-1}) = \{q_{j_1,i-1}, \dots, q_{j_k, i-1}\}$ (line 7); (2) generate $m-k$ queries by using $g(S_i, \mathcal{Q}) = \{q_{1,i}', \dots, q_{m-k,i}'\}$ (line 8); and (3) merge the queries from (1) and (2) to obtain $S_i = \{q_{j_1,i-1}, \dots, q_{j_k, i-1},$ $q_{1,i}', \dots, q_{m-k,i}'\}$ (line 9). Finally, we return the one with the highest fitness value (lines 11-12).

\begin{figure*}[t]
    \centering
    \includegraphics[width=0.72\linewidth]{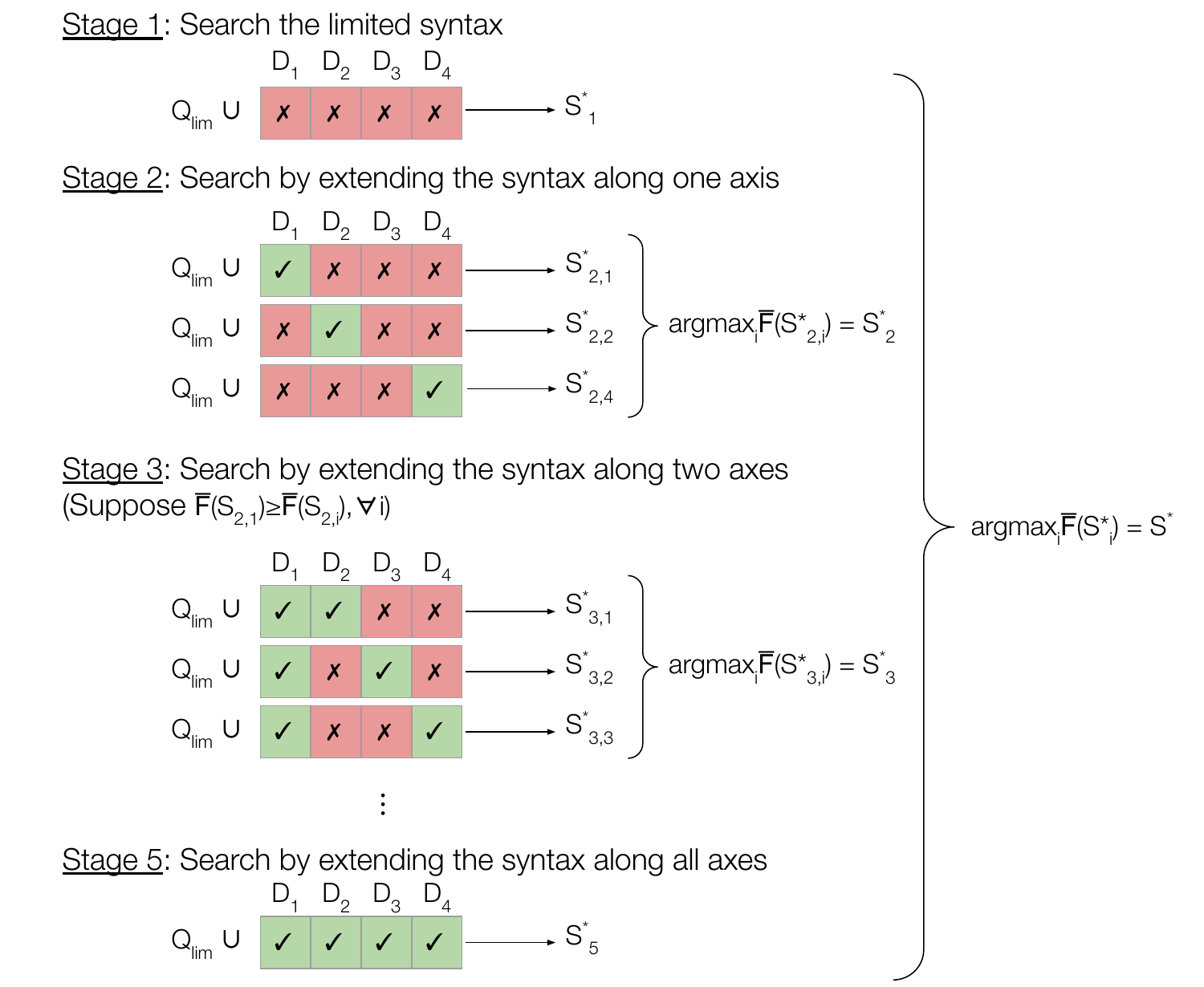}
    \caption{Illustration of QueryCheetah's multi-stage search.}
    \label{fig:method_full}
\end{figure*}

\textbf{Multi-stage search}. QueryCheetah performs multiple stages of local search. In each stage, it explores a different query syntax. In the stages, it extends the query syntax of the explored multisets along axes $D_1, \dots, D_4$. 

The multi-stage technique we use is similar to the Sequential Forward Feature Selection (SFFS)~\cite{chandrashekar2014survey} method in ML. It is a greedy algorithm solving the problem of finding an optimal subset of features for training an ML model. SFFS sequentially expands the considered subset of features in stages, by adding an unused feature, similar to how our technique expands the subset of syntax extensions.

Algorithm~\ref{algo:full} shows the pseudocode of the multi-stage search of QueryCheetah and Figure~\ref{fig:method_full} illustrates the stages.

First, in stage 0, we initialize the starting solution by using the generator $g$, $q_{j,1}=g(\emptyset, \mathcal{Q}_{lim}), j\in\{1,\dots,m\}$ and explore the limited syntax $\mathcal{Q}_{lim}$ (line 3). In stage $j$, $j\geq1$, we extend the limited syntax $\mathcal{Q}_{lim}$ along $j$ axes (lines 5-10). Denote by $D_j^*$ the subset with $j$ axes, $D_j^* \subseteq \{D_1, \dots, D_4\}, |D_j^*| = j$, in which we have discovered the solution $S_j^*$ with the highest fitness value in stage $j$ (line 11-12). Then, in stage $j+1$, we explore subsets $D_{j,i} \subseteq \{D_1, \dots, D_4\}$ that are supersets of $D_{j-1}^*$, $D_{j,i} \supset D_{j-1}^*$ (lines 13-14).  Denote by $S_{j,i}^*$ the corresponding multiset with the highest fitness discovered when exploring the extension $D_{j,i}$. Next, we choose the best extension in stage $j$ by taking the solution with the highest fitness $S_j^* = argmax_i\bar{F}(S_{j,i}^*)$. We initialize the search in stage $j+1$ by $S_j^*$ as the initial solution, and we iterate over it by applying the method $M$. Finally, we take the solution $S^*$ to be the one from the stage with the highest fitness, $s^* = argmax_j \bar{F}(S_j^*)$ (lines 16-17).

Note that the axes are not independent, i.e., the syntax can be extended along the axis $D_2$ only once it has been extended along $D_1$. Thus, the sets $D_{j,i}$ contain $D_2$, $D_2 \in D_{j,i}$ if and only if $D_{j-1}^*$ contains $D_1 \in D_{j-1}^*$ (lines 6-7).

\begin{algorithm}[t]

\SetCommentSty{mycommfont}
\SetAlgoNoLine

\SetKwInput{KwInput}{Input}                
\SetKwInput{KwOutput}{Output}              
\DontPrintSemicolon
  
  \KwInput{Query syntax $\mathcal{Q}_{lim}^a \subset \mathcal{Q}$, \;
  \hphantom{Input:  } Query syntax extension directions $\{D_1, \dots, D_e\}$, \;
  \hphantom{Input:  } Query multiset size $m$, \;
  \hphantom{Input:  } Number of attributes $n$, \;
  \hphantom{Input:  } Attributes $A$, \;
  \hphantom{Input:  } Target user record $r_u^{A'}$, \;
  \hphantom{Input:  } Number of iterations $num\_iters$, \;
  \hphantom{Input:  } Number of queries to keep $k$, \;
  \hphantom{Input:  } Solution fitness function $\bar{F}$, \;
  \hphantom{Input:  } Query importance function $\bar{f}$}
  \KwOutput{Multiset of queries}

  $D$ $\gets$ $\{D_1, \dots, D_e\}$ \tcp*{unexplored directions}
  $D'$ $\gets$ $\emptyset$ \tcp*{explored directions}
  $S^{0*}$ $\gets$ $\textsc{QueryCheetahSyntaxSubsetSearch}(\mathcal{Q}_{lim}^a, \emptyset)$ \;
  \tcc{For conciseness, we leave out the parameters $m, num\_iters, k, \bar{F}, \bar{f}$}
  \For{$i=1$ to e} {
    \For{$D_j$ in $D$} {
    \If{$D_j$ = $D_2$ \textbf{and} $D_1$ $\notin$ $D'$} {
        \textbf{continue}
    }
    
    $S^{i*}_j$ $\gets$ $\textsc{QueryCheetahSyntaxSubsetSearch}(\mathcal{Q}_{lim}^a$ $\cup D' \cup D_j, S^{(i-1)*})$ \tcp*{evaluate unexplored direction}
    }
    $r$ $\gets$ $argmax_j{\bar{F}(S^{i*}_j)}$  \tcp*{get the best unexplored direction}
    $S^{i*}$ $\gets$ $S^{i*}_j$ \tcp*{store the best query multiset in this stage}
    $D$ $\gets$ $D \setminus D_r$ \tcp*{remove the best direction from the unused}
    $D'$ $\gets$ $D' \cup D_r$ \tcp*{add the best to the used directions}
  }
  $S^{best\_stage*}$ $\gets$ $argmax_i(\bar{F}(S^{i*}))$ \; \tcp*{return the best multiset of all stages}
  \Return $S^{best\_stage*}$
\caption{\textsc{QueryCheetah}}
\label{algo:full}

\end{algorithm}

\section{Experimental setup}

\subsection{Datasets}\label{sec:datasets}
We evaluate QueryCheetah on three publicly available datasets.

Adult \cite{misc_adult_2} is a dataset extracted from the 1994 U.S. Census data. It contains $48,482$ records with $14$ socio-demographic attributes. The attributes include age, level of education, gender, and marital status. One of the $14$ attributes is a sensitive binary attribute describing whether the income of the individual is larger than \$50,000 (or not).

Census \cite{misc_census-income_(kdd)_117} contains $299,285$ records with $41$ socio-demographic attributes from the U.S. Census Bureau's Current Population Surveys. Similarly to Adult, it has one sensitive binary attribute for income.

Insurance \cite{misc_insurance5} contains car insurance $9,822$ records, each with $86$ demographic attributes aggregated over zip-codes, and one sensitive binary attribute about the interest in the insurance.

Adult and Census have missing values for some of their categorical columns. We have replaced the missing values with a special "Unknown" category. None of the three datasets have missing non-categorical values.

Adult and Census have $8$ and $33$ categorical attributes, respectively, while Insurance has none, except for the sensitive attribute. We map the categories of the categorical attributes to integers $0, 1, \dots$ for simplicity. For example, the values of a gender attribute $Female$ and $Male$ are mapped to $0$ and $1$.

\subsection{Evaluation metrics}

\textbf{Accuracy}.
We use the accuracy metric, denoted as $F$, obtained by playing the privacy game to compare the strength of the attacks discovered by QueryCheetah and competitor methods. 
Note that all methods are compared by using the accuracy metric $F$, despite that QuerySnout and QueryCheetah optimize the fitness function $\bar{F}$ in their search.

\textbf{Execution time}. We compare the execution time of the methods using Python's \textit{time} library on a Linux Virtual machine with a 40-core Intel Core processor and 108GB RAM.

\subsection{Attack parameters}

\textbf{Common parameters with state-of-the-art}. 
We use $n=6$ attributes, one sensitive and $5$ non-sensitive. We perform two sets of experiments, differentiated by how the non-sensitive attributes, $A'$, are selected. In the first set, we do not differentiate the type of the non-sensitive attribute and we sample without replacement $5$ non-sensitive attributes. In the second set, we differentiate the types of non-sensitive attributes and we sample without replacement $2$ categorical and $3$ non-categorical attributes. Differentiating the attributes depending on their type helps to evaluate the syntax extension along the axis $D_1$ that allows using \textit{BETWEEN}, but only with non-categorical attributes.

We do $5$ independent repetitions of randomly sampling the $n-1$ non-sensitive attributes. We attack users $u$ who are unique in these attributes $\forall v, v\neq u, r_v^{A'} \neq r_u^{A'}$. Evaluating on different attribute subsets amounts to considering different auxiliary information about the targeted individuals. For every user and repetition, we search for multisets of $m=100$ queries.

We repeat the privacy game $R=500$ times described in Section~\ref{subsec:game}. We calculate the fitness of the multisets by creating $f=3,000$ training and $g=1,000$ validation datasets of $8,000$ records randomly sampled from Adult and Census, and $1,000$ from Insurance (due to the total number of records in Insurance).

\textbf{QueryCheetah-specific parameters}. We generate $k=1$ new query and retain $m-1$ queries from the previous iteration $i-1$ when constructing the solution $S_i$ in iteration $i  (i>1)$. For the single-stage experiments that focus only on multisets in the limited syntax $\mathcal{Q}_{lim}$, we use $I=5,000$ total iterations. For the multi-stage experiments, we iterate $5,000 = I_1 = \dots = I_b$ iterations for every subset syntax (including the subset corresponding to the limited syntax). 

\textbf{State-of-the-art-specific parameters}.
For the experiments with QuerySnout, we use the recommended values of $200$ generations of the evolutionary search, each with a population of $|P| = 100$ solutions. We also use the recommended values for the evolutionary search probability parameters, such as $P_e$.

\section{Results}

\subsection{Attacking the limited syntax}\label{sec:main_results}

\begin{figure}
    \centering
    \includegraphics[width=1\linewidth]{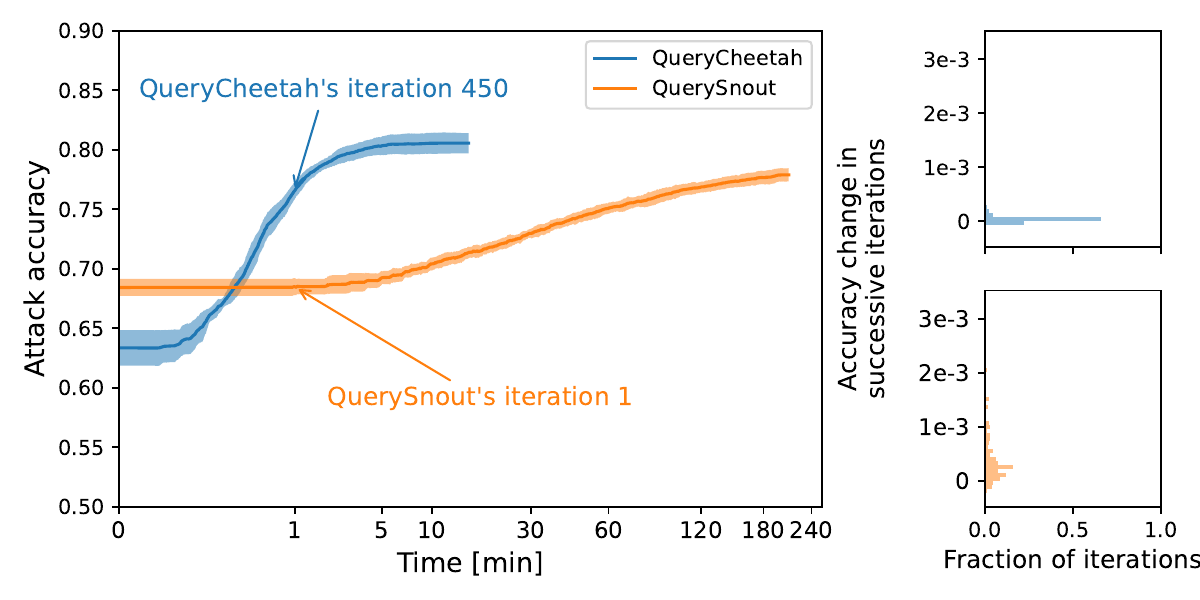}
    \caption{Execution time for attacking one target user in one repetition under the limited query syntax $\mathcal{Q}_{lim}$. We show the mean $\pm$ standard deviation of the test accuracy obtained from playing the privacy game on the Adult dataset using 5 repetitions, each repetition with 100 randomly selected users. On the right-hand side, we also show the accuracy change of the average attack from one iteration to the next.}
    \label{fig:main-adult}
\end{figure}

We first compare the performances of fully-automated methods, QueryCheetah (ours) and QuerySnout~\cite{cretu2022querysnout}, on the limited syntax $\mathcal{Q}_{lim}$. We track the performance of the best attacks they discovered up to every point in their execution time. 

Figure~\ref{fig:main-adult} shows the test accuracy of the best attacks discovered by the fully-automated methods on the Adult dataset. We here analyze and discuss the results on the Adult dataset, but the same conclusions hold for the Census and Insurance datasets.
Figure~\ref{fig:main-census-and-insurance} in  Appendix~\ref{sec:appendix_accuracy_over_time} shows the performance of the fully-automated methods on the Census and Insurance datasets.

There are three phases of the execution of the methods:
\begin{enumerate}
    \item In the first phase, QuerySnout's best discovered attacks have stronger inference capabilities than the best attacks discovered by QueryCheetah at that point in time. Before starting the search, QueryCheetah randomly samples and evaluates one candidate solution, while QuerySnout randomly samples a population of 100 solutions. Thus, at the start, QuerySnout has the upper hand as it has seen many more solutions, out of which it is likely that some would outperform QueryCheetah's initialization.
    \item In the second phase, QueryCheetah discovers attacks with stronger inference capabilities than QuerySnout's. QueryCheetah surpasses QuerySnout because of QueryCheetah's fast iterations. By the time QuerySnout has evaluated the population in iteration 1, QueryCheetah has performed 450 iterations and has discovered solutions that significantly outperform the ones considered by QuerySnout. 
    \item In the third and final phase, the performance of attacks discovered by QueryCheetah stabilizes and QueryCheetah terminates. At the same time, QuerySnout begins improving the performance of the discovered attacks. 
\end{enumerate}

Both methods improve over time by discovering stronger attacks, but observe different rates of improvement. 
QuerySnout performs fewer but longer iterations and shows better improvement from one iteration to the next. QueryCheetah takes 25 times more iterations, but each iteration is 450 times shorter on average. This leads to QueryCheetah being 18 times faster. The performance of attacks discovered by QueryCheetah improves the most in the second phase, while QuerySnout improves the most in the third phase, hours after QueryCheetah has terminated.

QueryCheetah discovers stronger attacks than the fully-automated method QuerySnout and than the semi-automated attack. Table~\ref{tab:limited_syntax} compares the test accuracies of the attacks discovered by QueryCheetah with the accuracies of the state-of-the-art attacks. QueryCheetah's attacks consistently outperform, or at worst case perform on par with the attacks discovered by the state-of-the-art methods across all three considered datasets.
Similarly to QuerySnout, QueryCheetah finds better attacks for the Insurance dataset than for Adult and Census. Insurance consists of all ordinal attributes, except for the sensitive one, which might help attacks to better isolate the target user.
 
\begin{table}
    \centering
    \begin{tabular}{lrrr}
          & Adult & Census & Insurance \\ \hline \hline
        QueryCheetah & $\mathbf{80.32\%}$ & $\mathbf{80.79\%} $  & $\mathbf{82.81\%} $ \\
        (fully-automated) & $\mathbf{\pm 1.43\%}$ & $\mathbf{\pm 1.09\%}$  & $\mathbf{\pm 1.06\%}$ \\ \hline
        QuerySnout & $77.77\% $ & $78.27\% $ & $80.14\%$ \\
		(fully-automated) & $\pm 0.52\%$ & $\pm 1.38\%$ & $ \pm 0.61\%$ \\ \hline
        Gadotti et al. & $76.34\%$ & $76.93\% $ & $73.02\% $ \\
        (semi-automated) & $ \pm 0.78\%$ & $\pm 1.38\%$ & $\pm 1.24\%$ \\
    \end{tabular}
    \caption{Comparison of attack accuracies discovered in the limited syntax $\mathcal{Q}_{lim}$ by the automated and the manual approaches. The comparison is performed over 5 repetitions on 100 users.}
    \label{tab:limited_syntax}
\end{table}

\subsection{Record-specific vulnerabilities} \label{sec:user_vulnerability}
The privacy vulnerabilities are record-specific. State-of-the-art methods evaluate the privacy loss of a QBS by attacking one user record at a time. Fully-automated methods require hours to attack one record in one repetition, as shown in Section~\ref{sec:main_results}. This limits the evaluation of privacy loss only to a highly limited number of records. Since vulnerabilities are record-specific, not testing for every record can, however, lead to missed vulnerabilities. Testing over multiple repetitions leads to a multiplicative factor of the time complexity. We here show how QueryCheetah, because of its efficiency, can be used to test the privacy leakage when attacking all target records from the three datasets, each record over $5$ repetitions. 

\begin{figure}
    \centering
    \includegraphics[width=0.8\linewidth]{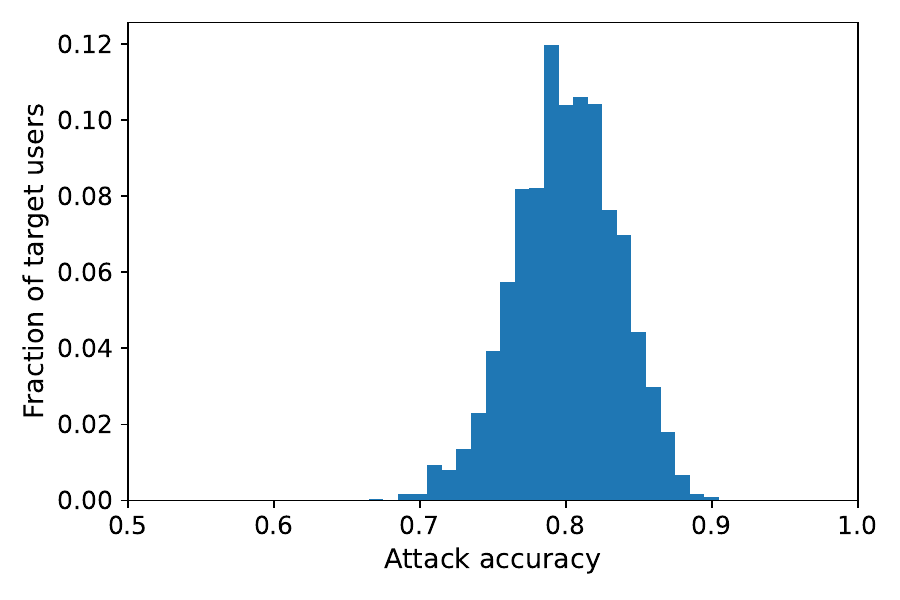}
    \caption{Histogram of the accuracies of attacks discovered by QueryCheetah on all unique users on the first repetition of Adult.}
    \label{fig:user_vulnerabilities_distribution}
\end{figure}

Figure~\ref{fig:user_vulnerabilities_distribution} shows the test accuracies of the discovered attacks of all target records in one repetition on Adult. Figure~\ref{fig:user_vulnerabilities_distribution_appendix} in Appendix~\ref{sec:appendix_vulnerabiltiies} shows the test accuracies of attacks across repetitions and across datasets. We here discuss one repetition of one dataset, but the same conclusions hold for all. QueryCheetah finds some records to be significantly more vulnerable than others. The strength of the discovered attacks varies by tens of percentage points. Thus, missing record-specific vulnerabilities can lead to seriously underestimating the privacy loss in practice. 
 
\subsection{Attacking the extended syntax}

\begin{table}
    \centering
    \begin{tabular}{lccc}
         & Adult & Census & Insurance \\ \hline \hline
        \multirow{ 2}{*}{Limited syntax $\mathcal{Q}_{lim}$} & $80.32\% $ & $80.79\% $  & $82.81\% $ \\
          & $\pm 1.43\%$ & $\pm 1.09\%$  & $ \pm 1.06\%$ \\ \hline
         \multirow{ 2}{*}{Extended syntax $\mathcal{Q}_{ext}$} & $80.86\%$ & $81.56\%$ & $85.38\%$ \\
          & $\pm 1.15\%$ & $\pm 1.20\%$ & $\pm 0.37\%$ \\
    \end{tabular}
    \caption{Accuracies of attacks discovered by QueryCheetah over 5 repetitions, each on 100 users.}
    \label{tab:full_syntax}
\end{table}

Evaluating the privacy guarantees of a QBS that supports a rich query syntax is a difficult problem. Supporting a richer query syntax leads to a larger attack surface. The possible number of attacks in a richer syntax can be exponentially higher. As we have shown in Table~\ref{tab:sizes}, searching for attacks in the extended syntax $\mathcal{Q}_{ext}$ leads to searching a vast search space. This highlights the need for an efficient search method.

QueryCheetah's speed allows it to explore vast search spaces. We instantiate it for AIAs under the extended syntax $\mathcal{Q}_{ext}$. 

Table~\ref{tab:full_syntax} presents the performance of the attacks discovered under the extended syntax and compares it with the performance of attacks in the limited syntax. In the extended syntax, QueryCheetah has discovered better-performing attacks than in the limited one for Insurance, and marginally better attacks for Adult and Census. The performance gap of attacks for Insurance and for Adult and Census increases when extending the syntax; attacks for Insurance can also make use of extensions along a axis that can only be used for ordinal attributes, i.e., using ranges as conditions.

Table~\ref{tab:extended_syntax_appendix} in Appendix~\ref{sec:appendix_full_syntax} presents results of the analogous experiments where we select $3$ out of the $n=5$ attributes for Adult and Census to be ordinal. In this scenario, the discovered attacks against both datasets in the extended syntax outperform the ones in the limited syntax. They also reduce the performance gap with the attacks for Insurance.

\subsection{Mitigations}
Diffix has implemented additional defenses, which we refer to as mitigations, to thwart discovered attacks.
We here analyze the effectiveness of Diffix's mitigations (described in Section~\ref{sec:mitigations_diffix}) in two ways. First, we apply them post-hoc to the attacks discovered in the limited $\mathcal{Q}_{lim}$ and the extended $\mathcal{Q}_{ext}$ syntax. Second, we instantiate QueryCheetah to search for attacks against a target QBS with these mitigations implemented, allowing it to discover workarounds.

Table~\ref{tab:mitigations} presents the accuracy of the attacks using both approaches. First, deploying the mitigation post-hoc against the discovered attacks significantly hinders the attacks' performance.  Second, searching for attacks against a QBS that implements the mitigations, helps an attacker to find stronger attacks. The effectiveness of the mitigations is pronounced for Adult and Census. The conclusions of both steps hold for attacks in both the limited $\mathcal{Q}_{lim}$ and the extended syntax $\mathcal{Q}_{ext}$.

The standard deviation of the attacks' test accuracy significantly increases after deploying the mitigations in both steps. This signals that the mitigations protect some attributes more than others. However, the second step, where we search for attacks against QBSs that implement the mitigations, can help in finding workarounds.

Table~\ref{tab:mitigations_appendix} in Appendix~\ref{sec:appendix_mitigations} presents the analogous results in the case when $3$ out of the $n=5$ attributes are ordinal, for which the same conclusions hold.

\begin{table*}
    \centering
    \begin{tabular}{llrrr}
        & & Adult  & Census & Insurance \\\hline
        \multirow{3}{*}{Limited syntax  $\mathcal{Q}_{lim}$} & \multirow{1}{*}{No mitigations} &  $80.32\% $  $\pm 1.43\%$ & $80.79\% $ $\pm 1.09\%$ & $82.81\% $ $ \pm 1.06\%$ \\
        \cline{2-5}
         & \multirow{1}{*}{Post-hoc mitigations} & $72.70\%$ $\pm 5.76\%$ & $73.63\%$ $\pm 7.19\%$  & $81.52\%$ $\pm 1.18\%$ \\
         \cline{2-5}
         & Including mitigations in the search  & $74.72\%$ $\pm 5.13\%$ & $75.55\%$ $\pm 5.99\%$ & $81.55\%$ $\pm 1.20\%$ \\  
         \hline 
         \multirow{3}{*}{Extended syntax  $\mathcal{Q}_{ext}$} &  \multirow{1}{*}{No mitigations} &  $80.86\%$ $\pm 1.15\%$ & $81.56\%$ $\pm 1.20\%$ & $85.38\%$ $\pm 0.37\%$ \\ 
         \cline{2-5}
         & \multirow{1}{*}{Post-hoc mitigations} & $72.21\%$ $\pm 6.00\%$ & $72.96\%$ $\pm 7.70\%$ & $82.45\%$ $\pm 0.82\%$ \\ 
         \cline{2-5}
         & Including mitigations in the search & $75.83\%$ $\pm 5.16\%$ & $75.87\%$ $\pm 5.78\%$ & $84.13\%$ $\pm 0.44\%$ \\
    \end{tabular}
    \caption{Impact of mitigation on attacks discovered by QueryCheetah over 5 repetitions, each on 100 users.}
    \label{tab:mitigations}
\end{table*}

\section{Discussion}

\subsection{Analysis of the discovered attacks}
We have shown QueryCheetah to discover stronger attacks than state-of-the-art~\cite{cretu2022querysnout}. We here analyze the attacks discovered by QueryCheetah to gain insights into the vulnerability(/ies) they exploit and compare them to QuerySnout. More specifically, we isolate all the queries of a given type, for example difference-like queries, from the discovered attacks and report the proportion of the attack's success they account for. The procedure is described in more detail in Appendix~\ref{sec:appendix_analysis_of_attacks}. In line with QuerySnout~\cite{cretu2022querysnout}, we do not verify the uniqueness condition for every target user, test dataset and difference query pair, for computational reasons. Instead, we study difference-like queries, queries that have the same syntax as difference queries (cf Equation~(\ref{eq:difference_queries})) but do not necessarily fulfill the uniqueness condition that the target user $u$ is the only user in the userset of $q_2$, $u \in Y(D, q_2)$, who is not in the userset of $q_1$, $u \notin Y(D, q_2)$ for all test datasets $D \in \{D_1^{test}, \ldots, D_{R}^{test}\}$.

Table~\ref{tab:attack_explainability_adult_restricted_syntax} shows that in the limited syntax $\mathcal{Q}_{lim}$, QueryCheetah exploits essentially the same vulnerability as QuerySnout: difference-like queries (DFLQ). DFLQ indeed account for $98\%$ of QueryCheetah's attack accuracy and 97\% of QuerySnout's~\cite{cretu2022querysnout}. Our results further show that the performance gap between QueryCheetah and QuerySnout is likely to come from the number of discovered difference queries, with QueryCheetah discovering more difference-like queries than QuerySnout ($70$ vs $30$ when queries are counted with multiplicities and $40$ vs $25$ unique queries). We obtain similar results on the  Census and Insurance datasets (Tables \ref{tab:attack_explainability_census_restricted_syntax} and \ref{tab:attack_explainability_insurance_restricted_syntax}).

\begin{table}
    \centering
    \begin{tabular}{lrr}
          & QueryCheetah & QuerySnout \\ \hline \hline
        Number of queries & $71.05 \pm 6.82$ & $28.98 $ $\pm 6.62$ \\
        Number of unique queries & $43.20 \pm 1.20$ & $26.32 $ $\pm 5.80$ \\
        Attack accuracy using & \multirow{2}{*}{$79.43\% \pm 1.62\%$} & \multirow{2}{*}{$76.25\% $ $\pm 0.72\%$} \\
        the subset of queries & & \\
        Percentage of accuracy  & \multirow{2}{*}{$98.89\% \pm 0.30\%$} & \multirow{2}{*}{$98.01\% \pm 0.37\%$}  \\
        accounted for by the subset & & \\
    \end{tabular}
    \caption{Limited syntax $\mathcal{Q}_{lim}$: Performance when isolating all the difference-like queries from the discovered attacks on the Adult dataset. The comparison is performed over 5 repetitions on 100 users.}
    \label{tab:attack_explainability_adult_restricted_syntax}
\end{table}

Table~\ref{tab:attack_explainability_adult_extended_syntax} (left) shows that in the extended syntax $\mathcal{Q}_{ext}$ the difference-like queries are not the only vulnerability exploited by QueryCheetah. The discovered attacks on Adult contain indeed only $27$ difference-like out of the $100$ queries ($17$ unique on average) and, taken together, they only account for 83\% of the accuracy of the attack. 

We thus introduce the notion of generalized difference-like queries. These extent the notion of difference-like queries to the extended syntax $\mathcal{Q}_{ext}$, such that (a) instead of $a_i = r_u^i$, $i\in\{1,\dots,n\}$, we allow any condition on the attribute $a_i$ that selects the target user $u$, and (b) instead of $a_i \neq r_u^i$, $i\in\{1,\dots,n\}$, we allow any condition that excludes the target user $u$. Note that, by definition, generalized difference-like queries are a superset of difference-like queries.

Table~\ref{tab:attack_explainability_adult_extended_syntax} (right) shows that generalized difference-like queries account for $97\%$ of the attack accuracy on the Adult  dataset with similar results on Census and Insurance
(Tables~\ref{tab:attack_explainability_census_extended_syntax} and \ref{tab:attack_explainability_insurance_extended_syntax}).

\begin{table}
    \centering
    \begin{tabular}{lrr}
    
          & DFLQ & GDFLQ \\\hline \hline
        Number of queries & $27.21 \pm 3.99$ & $58.34 $ $\pm 6.30$ \\
        Number of unique queries & $16.95 \pm 1.79$ & $37.28 $ $\pm 1.55$ \\
        Attack accuracy using & \multirow{2}{*}{$67.34\% \pm 1.15\%$} & \multirow{2}{*}{$78.94\% $ $\pm 1.51\%$} \\
        the subset of queries & & \\
        Percentage of accuracy & \multirow{2}{*}{$83.37\% \pm 1.74\%$} & \multirow{2}{*}{$97.62\% \pm 0.62\% $}  \\
        accounted for by the subset & & \\
    \end{tabular}
    \caption{Extended syntax $\mathcal{Q}_{ext}$: Performance when isolating all difference-like (DFLQ) and generalized difference-like queries (GDFLQ) from the discovered attacks on the Adult dataset. The comparison is over 5 repetitions on 100 users.}
    \label{tab:attack_explainability_adult_extended_syntax}
\end{table}

\subsection{Asymptotic time complexity} 
We now analyze the asymptotic complexity of QueryCheetah. The attacks are user record-specific and QueryCheetah searches for attacks one record at a time. For each record, it uses a local search technique over $I$ iterations. In each iteration, it (1) estimates the fitness of the multiset at that iteration with $f+g$ datasets and repeats the privacy game $R$ times, each time with a different shadow dataset, (2) chooses $k$ queries to retain, and generates $m-k$ new queries. The asymptotic complexity of QueryCheetah's search for a given target user is thus $O(I\cdot(m\cdot(f+g+R) +  m\cdot n\cdot(m-k)))$. Retaining $m-1$ queries at each iteration leads us to a final asymptotic complexity of $O(I\cdot m\cdot(f+g+R + n)))$. Note that this is a factor $P$ faster than QuerySnout's asymptotic time complexity, $O(I\cdot P \cdot m\cdot(f+g+R + n)))$, where $P$ is the number of multisets in the population.

\subsection{Stability before termination}
To analyze the search procedure's stability near termination, we compare the fitness at the last iteration, $\bar{F}_{I}$, with the mean fitness value in the last 100 iterations $\mu = (\bar{F}_{I-100} + \dots \bar{F}_{I-1}, \bar{F}_{I})$, where $\bar{F}_i$ denotes the fitness of the multiset at iteration $i$, $i\in \{1, \dots, I\}$. Table~\ref{tab:stability} shows $|\bar{F}_{I} - \mu|$ to be systematically well below the standard deviation of the attack accuracies, suggesting that the search method is stable near termination. 

\begin{table}
    \centering
    \begin{tabular}{llccc}
         & & Adult & Census & Insurance \\ \hline \hline
         & \multirow{2}{*}{No mitigations} & $0.17\% $ & $0.13\% $  & $0.15\% $ \\
         \multirow{1}{*}{Limited } & & $\pm 0.05\%$ & $\pm 0.03\%$  & $ \pm 0.02\%$ \\ \cline{2-5}
        \multirow{1}{*}{syntax $\mathcal{Q}_{lim}$} & \multirow{2}{*}{With mitigations} & $0.19\% $ & $0.14\% $  & $0.16\% $ \\
          & & $\pm 0.03\%$ & $\pm 0.02\%$  & $ \pm 0.04\%$ \\ \hline
          & \multirow{2}{*}{No mitigations} & $0.19\%$ & $0.16\%$ & $0.18\%$ \\
          \multirow{1}{*}{Extended }& & $\pm 0.04\%$ & $\pm 0.03\%$ & $\pm 0.03\%$ \\ \cline{2-5}
          \multirow{1}{*}{syntax $\mathcal{Q}_{ext}$} & \multirow{2}{*}{With mitigations} & $0.14\%$ & $0.13\%$ & $0.15\%$ \\
          & & $\pm 0.02\%$ & $\pm 0.02\%$ & $\pm 0.03\%$ \\
    \end{tabular}
    \caption{Stability of fitness near termination measured by $|\bar{F}_{I} - \mu|$ over 5 repetitions, each on 100 users.}
    \label{tab:stability}
\end{table}

\subsection{Extension to multi-value sensitive attribute}
We have so far assumed, for simplicity, the sensitive attribute $a_n$ to be binary. QueryCheetah, however, can also be instantiated against multi-value sensitive attributes with only a minor change: the privacy game in Section~\ref{subsec:game} needs to be modified to allow the defender Alice to use any of the possible values for the sensitive attributes. Namely, in her second step, instead of sampling from $Uniform(\{0,1\})$ (equivalent with $Bernoulli(0.5)$), she would sample from $Uniform(\{\mathcal{V}_n\})$, where $\mathcal{V}_n$ denotes the set of possible values of the sensitive attribute $a_n$. 

We however expect that more shadow datasets would likely be needed to achieve optimal performance for multi-valued attributes. The literature suggests keeping the number of examples per possible target value constant~\cite{abramovich2019classification}, meaning a roughly linear increase of the number of shadow datasets with respect to the increase of the number of possible values of the sensitive attribute.

\subsection{Generalizability to other QBSs}
We here describe how QueryCheetah can be applied to other QBSs.

First, QueryCheetah can be instantiated out-of-the-box against QBSs supporting the same syntax we have focused on in this work ($\mathcal{Q}_{lim}$ and $\mathcal{Q}_{ext}$), regardless of the defense mechanisms used. This is because QueryCheetah is agnostic to the defense mechanism.

Second, we argue that adapting QueryCheetah to discover attacks against QBSs that support a different query syntax should not be difficult in general as it would only require two modifications: (1) modifying the method for generating queries $g(S_i, \mathcal{Q})$ in the new query syntax, and (2) defining axes of extension for the new syntax, similarly to the axes $D_1, \dots, D_4$ defined in Section~\ref{sec:diffx_query_syntax}.

Third, QueryCheetah can also be adapted to QBSs providing differential privacy guarantees~\cite{dwork2006calibrating}. This requires defining a strategy for dividing the privacy budget between the queries asked. We discuss in Appendix~\ref{sec:appendix_dp_qbss} several such possible strategies.

\subsection{Extension to membership inference attacks}
Membership inference attacks (MIAs) are an important and popular empirical metric of privacy risk, allowing e.g., to audit the guarantees of formal protections such as differential privacy~\cite{jagielski2020auditing}. We here show how QueryCheetah can be instantiated to discover MIAs with minimal modifications to the privacy game (Section~\ref{subsec:game}): 

First, Alice would not always add the target user's record $r_u^{A'}$ with a sampled value of the sensitive attribute $b_j$, $j\in\{1,\dots,s_D\}$ into $D_i^{test}$. Instead, she would first sample a value $b'\sim Bernoulli(0.5)$ and add the target user's record if and only if $b'$ is $1, b'=1$. Otherwise, she would sample and add another record from $\mathcal{D}$.

Second, Alice would not need to break the correlation between the sensitive attribute and the other attributes, since MIAs focus on the record-level and not the attribute-level aims like AIAs. 

Third, Bob's goal would be to predict the membership of the target user $u$, i.e., predict $b'$. As before, this is a binary classification task where the baseline is a random coin flip with $50\%$ accuracy.

Please note that these first two modifications lead to slightly different datasets than those used for AIAs.

Table~\ref{tab:mias} shows that the MIAs discovered by QueryCheetah against Diffix achieve out-of-the-box an accuracy of $77\%$ in the limited syntax and $79\%$ in the extended syntax on average across datasets, substantially outperforming a random guess (0.5) in both syntaxes. 

\begin{table}
    \centering
    \begin{tabular}{lccc}
         & Adult & Census & Insurance \\ \hline \hline
        \multirow{ 2}{*}{Limited syntax $\mathcal{Q}_{lim}$} & $74.48\% $ & $78.54\% $  & $78.15\% $ \\
          & $\pm 1.26\%$ & $\pm 0.83\%$  & $ \pm 0.63\%$ \\ \hline
         \multirow{ 2}{*}{Extended syntax $\mathcal{Q}_{ext}$} & $75.03\%$ & $79.29\%$ & $81.27\%$ \\
          & $\pm 1.27\%$ & $\pm 0.78\%$ & $\pm 1.75\%$ \\
    \end{tabular}
    \caption{Accuracies of membership inference attacks discovered by QueryCheetah over 5 repetitions, each on 100 users.}
    \label{tab:mias}
\end{table}

\subsection{Impact of correlations between non-sensitive attributes}\label{sec:correlations}
In our evaluation, we intentionally broke the correlations between the sensitive and the non-sensitive attributes in the privacy game. This ensures that we measure the pure success of our attack and not the ability of a classifier to infer the value of the sensitive attribute based on the non-sensitive ones (imputation).

We here investigate the performance of QueryCheetah on datasets with little to no correlations between the non-sensitive attributes by (1) generating synthetic datasets by independently sampling values from one-way marginals, and (2) instantiating QueryCheetah on these synthetic datasets. One could e.g. imagine an attack to leverage a strong correlation between two attributes by replacing one attribute with another in a query to generate another answer with a different noise sample.

More specifically, we generate synthetic datasets that have the same attributes $a_1, \dots, a_n$, and the same possible values for them $\mathcal{V}_1, \dots, \mathcal{V}_n$ as Adult, Census, and Insurance. We also generate the same number of synthetic records by independently sampling values from $\mathcal{V}_i$ for each attribute $a_i$, $i\in\{1,\dots,n\}$. Independently sampling the values ensures that by design the synthetic datasets have little to no correlations. 

Table~\ref{tab:synthetic} shows how the accuracies we obtain against the synthetic datasets, where there are no correlations between the non-sensitive attributes, do not substantially differ from the accuracies we obtain against the original datasets (Table~\ref{tab:full_syntax}).

\begin{table}
    \centering
    \begin{tabular}{lccc}
         & Synthetic & Synthetic & Synthetic \\
         & Adult & Census & Insurance \\ \hline \hline
        \multirow{ 2}{*}{Limited syntax $\mathcal{Q}_{lim}$} & $79.72\% $ & $81.37\% $  & $81.86\% $ \\
          & $\pm 0.84\%$ & $\pm 0.43\%$  & $ \pm 1.00\%$ \\ \hline
         \multirow{ 2}{*}{Extended syntax $\mathcal{Q}_{ext}$} & $80.29\%$ & $81.96\%$ & $83.39\%$ \\
          & $\pm 0.78\%$ & $\pm 0.73\%$ & $\pm 0.51\%$ \\
    \end{tabular}
    \caption{Accuracies of attacks discovered by QueryCheetah on synthetic datasets with no correlations between non-sensitive attributes over 5 repetitions, each on 100 users.}
    \label{tab:synthetic}
\end{table}

\section{Related work} \label{sec:related_work}
\subsection{Attacks against QBSs}
A rich body of research has explored adversarial attacks against QBSs in general. In 1979, Denning et al.~\cite{denning1979tracker} proposed multisets of queries that constitute attacks against a QBS that implements only a bucket suppression mechanism. In 2003, Dinur et al.~\cite{dinur2003revealing} proposed a reconstruction attack against a QBS that answers $nlog^2n$ queries by perturbing each answer with noise of at most $O(\sqrt{n})$, such that each query selects a random subsets of users in the dataset $Y(D,q) \subseteq U$. Subsequent work has improved the attack’s robustness to distortion~\cite{dwork2007price} and enhanced it to work with fewer queries~\cite{dwork2008new}. Chipperfield et al.~\cite{chipperfield2016australian} and Rinnot et al.~\cite{rinott2018confidentiality} proposed manual AIAs against a QBS that implements bucket suppression and bounded seeded noise addition. Pyrgelis et al.~\cite{pyrgelis2017knock} proposed an MIA that uses counting query answers about the number of people in a certain area at a given time. They automate the inference by training an ML classifier on shadow datasets, a technique later adapted by QuerySnout (explained in Section~\ref{sec:querysnout}) and which we also use. 

\subsection{Attacks against Diffix}
Among real-world QBSs, Diffix has received the most significant research attention focusing on attack methods. Three types of attacks have been discovered: AIA, MIA, and reconstruction attack. Gadotti et al.~\cite{gadotti2019signal} proposed an AIA against Diffix (see Section~\ref{sec:gadotti}). Earlier versions of Diffix were found to be vulnerable to MIAs and reconstruction attacks. Pyrgelis et al.~\cite{benthamsgazeLocationTime} proposed an MIA on location data based on their earlier work~\cite{pyrgelis2017knock}. Their attack uses a multiset of tens of thousands of queries. Cohen et al.~\cite{cohen2018linear} and Joseph et al.\cite{differentialprivacyReconstructionAttacks} proposed reconstruction attacks, that also require a high number of queries, proportional to the dataset size $s_D$, based on the work of Dinur et al.~\cite{dinur2003revealing}. Both attacks against Diffix use the inference approach of Dinur et al.~\cite{dinur2003revealing} on manually identified class of query multisets that satisfy the condition of uniform random samples of users.   

\subsection{Automation of privacy attacks}

To the best of our knowledge, QuerySnout~\cite{cretu2022querysnout}, described in Section~\ref{sec:querysnout} is the only method for automated discovery of privacy attacks against general-purpose QBSs. The research field of automated attacks specific to DP violations has been much more prolific. The methods proposed in this line of work automate the inference or the search for neighboring datasets or output events. Wang et al.~\cite{wang2020checkdp} automatically infer a DP violation using a static program analysis tool that analyzes the software with the aim to verify correct implementation. Ding et al.~\cite{ding2018detecting} introduce a method called DPStat that relies on a statistical hypothesis test that detects a vulnerability of DP guarantees. Bischel et al.~\cite{bichsel2018dp} introduce DP-Finder that uses symbolic differentiation and gradient descent to find neighboring datasets and outputs that violate the DP guarantees. They have later introduced a method called DP-Sniper~\cite{bichsel2021dp}, which uses an automated approach based on an ML classifier to detect DP vulnerability, a method that also emphasizes the importance of fast automated methods.

Shokri et al.~\cite{shokri2017membership} have proposed a technique, called shadow modeling, used in automated attacks. They have used it to show that releasing an ML model can reveal the membership of a user in its training data. Adaptations of the shadow modeling technique have also been used to reconstruct data examples from the training set~\cite{balle2022reconstructing}, infer properties of the training set~\cite{ateniese2015hacking,ganju2018property}, and infer user membership in synthetic data~\cite{stadler2022synthetic}.

\subsection{Differential privacy as a defense}
Differential privacy~\cite{dwork2006calibrating} is a mathematically rigorous definition of a privacy guarantee which was proposed as a defensive solution to the reconstruction attack of Dinur et al.~\cite{dinur2003revealing}. It gives worst-case theoretical guarantees against a strong attacker. Implementations of DP in practice however can be difficult. First, they often use relatively large values for $\epsilon$ to obtain the desired utility, sometimes to an extent that undermines the intended theoretical guarantees~\cite{abowd2018us, rogers2020members}. Second, they provide relaxed DP guarantees, providing weaker, for example, event-level instead of user-level guarantees~\cite{houssiau2022difficulty, chetty2022social}. Third, as regular data releases are difficult with a bounded privacy budget, many DP deployments regularly (e.g., monthly) reset their budget~\cite{amazonDataCollaboration, rogers2020linkedin}, which can invalidate the guarantees in the long run. 

\section{Conclusion}
In this paper, we propose a novel method, QueryCheetah, for efficiently and automatically searching privacy vulnerabilities against query-based systems. We instantiate the method for discovering attribute inference attacks against a popular real-world QBS, Diffix.

First, we evaluate QueryCheetah's performance against existing state-of-the-art methods: outperforming existing semi- and fully-automated methods in the accuracy of discovered attacks against three datasets while being 18 times faster than fully-automated methods.
Second, we show how QueryCheetah can more thoroughly evaluate the privacy loss of a QBS by attacking all target users. Namely, vulnerabilities against a QBS are user-specific, and thus, computationally expensive methods that target a limited number of users can lead to missed vulnerabilities. Third, we evaluate QueryCheetah in a richer query syntax, which makes up for a vast search space of possible attacks. Fourth and final, we evaluate the effectiveness of defenses implemented to thwart attacks and show them to decrease the accuracy of attacks.

We show that it both outpaces and outperforms existing methods by discovering better-performing attacks in a shorter time period. Using the fast method, we then target all target users in the dataset. 

 
\textbf{Acknowledgments}    
This work has been partially supported by the CHEDDAR: Communications Hub for Empowering Distributed ClouD Computing Applications and Research funded by the UK EPSRC under grant numbers EP/Y037421/1 and EP/X040518/1 and by the PETRAS National Centre of Excellence for IoT Systems Cybersecurity, funded by the UK EPSRC under grant number EP/S035362/1.
We acknowledge computational resources provided by the Imperial College Research Computing Service.~\footnote{http://doi.org/10.14469/hpc/2232}
The authors would like to thank the anonymous reviewers and shepherd for their feedback.

\bibliographystyle{plain}
\bibliography{refs}

\appendix

\begin{algorithm} [h]
\SetCommentSty{mycommfont}
\SetAlgoNoLine
\SetKwInput{KwInput}{Input}                
\SetKwInput{KwOutput}{Output}              
\DontPrintSemicolon
  \KwInput{Query syntax $\mathcal{Q'}^a \subseteq \mathcal{Q}$, \;
  \hphantom{Input:  } Number of attributes $n$, \;
  \hphantom{Input:  } Attributes $A$, \;
  \hphantom{Input:  } Target user record $r_u^{A'}$ \;
  }
  \KwOutput{Random query}
  $conditions$ $\gets$ $[]$ \;
  
  \For{$i=0$ to n \tcp*{for each attribute}} {
  \tcp{Check which comparison operators are supported}
    $more\_complex\_operators$ $\gets$ $[]$ \; 
    $simpler\_operators$ $\gets$ $[=, \neq]$ \;
    $skip\_operator$ $\gets$ $\{\perp\}$
    \If{"BETWEEN" $\in \mathcal{O}$ \textbf{and} $is\_ordinal(a_i)$} {
        $more\_complex\_conditions.append("BETWEEN")$
    }
    \If{"IN" $\in \mathcal{O}$} {
        $more\_complex\_conditions.append("IN")$
    }
    \If{"NOT IN" $\in \mathcal{O}$} {
        $more\_complex\_conditions.append("NOT\;IN")$
    }
  \tcp{Sample a comparison operator}
    \eIf{more\_complex\_operators = $\emptyset$}{
        $operator$ $\gets$ $random\_sample(\{\perp, =, \neq\})$
    } {
        $type\_of\_operator$ $\gets$ $random\_sample$$(skip\_operator, simpler\_operators$, $more\_complex\_operators)$ \;
        $operator$ $\gets$ $random\_sample(type\_of\_operator)$
    }
        \tcp{Sample a value}
    $value$ $\gets$ $get\_value\_given\_comparison\_operator(operator)$ \tcp*{Algorithm~\ref{algo:random_value}}

    $condition$ $\gets$ $a_i$ + $operator$ + $value$\;
    $conditions.append(condition)$\;

    }
    \tcp{Combine all conditions into a query}
    \Return "SELECT count(*) WHERE" + join(conditions, "AND")  
\caption{\textsc{RandomQuery}}
\label{algo:random_query}
\end{algorithm}

\begin{algorithm}

\SetCommentSty{mycommfont}
\SetAlgoNoLine

\SetKwInput{KwInput}{Input}                
\SetKwInput{KwOutput}{Output}              
\DontPrintSemicolon
  
  \KwInput{Query syntax $\mathcal{Q'}^a \subseteq \mathcal{Q}$, \;
  \hphantom{Input:  } Number of attributes $n$, \;
  \hphantom{Input:  } Attributes $A$, \;
  \hphantom{Input:  } Target user record $r_u^{A'}$ \;
  }
  \KwOutput{Condition value}
  \eIf{only\_target\_user\_values\_supported($\mathcal{Q}'^a$)} {
  $value$ $\gets$ $r^i$
  }
  {
    $possible\_range\_width$ $\gets$ $[1\cdot10^{-2}, 2\cdot10^{-2}, 5\cdot10^{-2}, \dots, 1, 2, 5]$ \;
    $width$ $\gets$ $random\_sample(possible\_range\_width)$ \;
    $k_1$ $\gets$ $round(r^i / (2 \cdot width))$ \;
    $k_2$ $\gets$ $round((2 \cdot r^i - width) / (4 \cdot width))$ \;
    $offset_1$ $\gets$ $width \cdot 2\cdot k_1 $ \;
    $offset_2$ $\gets$ $width \cdot(2\cdot k_2 + 0.5)$ \;
    $offset$ $\gets$ $offset_1$ \textbf{if} $|r^i - offset_1| < |r^i - offset_2|$ \textbf{else} $offset_2$ \;

    $aux$ $\gets$ $sample\_uniform\_random(D_{aux}^i, a_i, r_i)$ \tcp*{sample a value from the auxiliary dataset different than the target user's value}
    \If{operator = "BETWEEN"} {$value \gets (offset, offset + width)$}
    \If{operator = "="} {$value \gets (r^i, aux)$}
    \If{operator = "$\neq$"} {$value \gets sample\_random(r^i, offset)$}
    \If{operator = "IN" \textbf{or} operator = "NOT IN"} {$value \gets (r^i, random\_sample(aux, -1000000))$}
  }

    \Return $value$
  
\caption{\textsc{RandomValueForComparisonOperator}}
\label{algo:random_value}
\end{algorithm}

\section{Generating random queries}\label{sec:random_query}
QueryCheetah uses a generator $g$ for generating random queries. Its pseudocode is described in Algorithm~\ref{algo:random_query}. It relies in one of its steps on a method for randomly generating values, which we describe in Algorithm~\ref{algo:random_value}.

Overall, we construct the filtering conditions in the \textit{WHERE} clause by iterating over the attributes $A$ (line 2). For each attribute $a_i$, we generate the comparison operator $c_i$ as follows.

First, we generate a comparison operator $c_i$. We check which comparison operators $\mathcal{C}$ are supported by the query syntax $\mathcal{Q}$ for that attribute (lines 4-14). If none of the comparison operators in extensions $D_2, D_3,$ and $D_4$ are supported, we default to the approach in the literature that randomly samples among the possibilities, i.e., $\mathcal{C} = \{\perp, =, \neq\}$ (line 16). Otherwise, we sample at uniform random which \textit{type} of an operator to include, a skip, i.e., $\perp$, a simple, i.e., $=, \neq$, or operators in the extensions $D_2,\dots,D_4$ (line 18), and sample uniformly at random from the operators of that type (line 19). 

Second, once we have generated the comparison operator $c_i$, we generate the value $v_i$ as per Algorithm~\ref{algo:random_value}.

Third and final, we concatenate the attribute $a_i$, the comparison operator $c_i$, and the value $v_i$ to form a filtering condition $a_i\;c_i\;v_i$ (line 22), and we concatenate all $n$ conditions with a logical operator $AND \in \mathcal{O}$ as a connector $a_1\;c_1\;v_i AND \dots a_n\;c_n\;v_n$ to create the query (line 25).

\section{Attacking the limited syntax}\label{sec:appendix_accuracy_over_time}
\begin{figure}
    \centering
    \begin{subfigure}[b]{\linewidth}
    \includegraphics[width=1\linewidth]{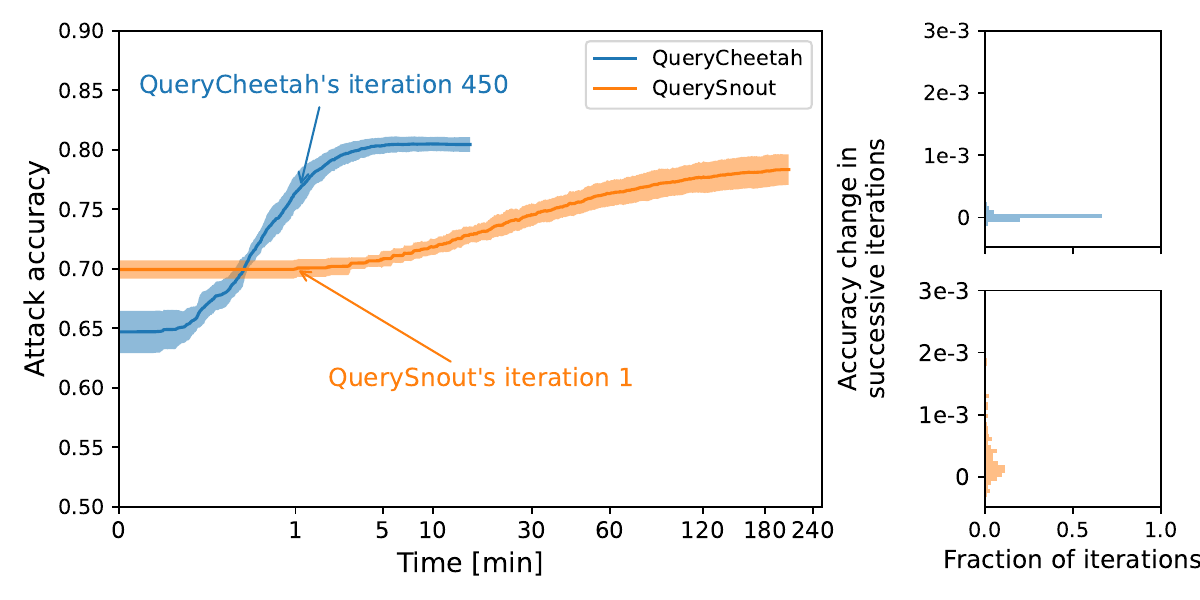}
    \caption{Census dataset}
    \label{subfig:main-census}
    \end{subfigure}
    \begin{subfigure}[b]{\linewidth}
    \includegraphics[width=1\linewidth]{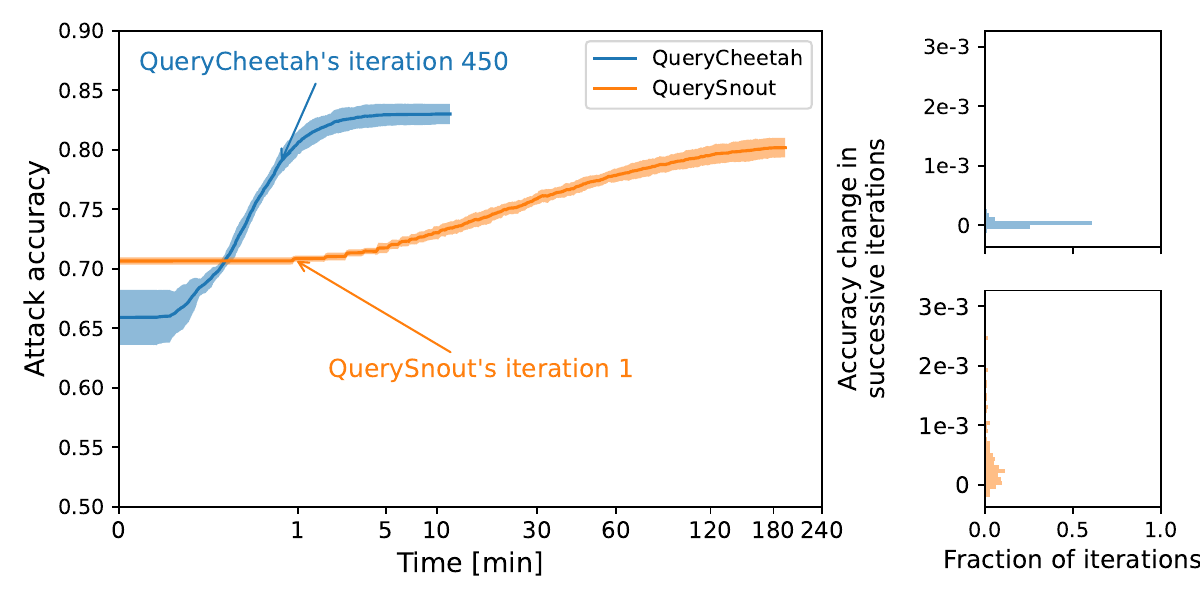}
    \caption{Insurance dataset}
    \label{subfig:main-insurance}
    \end{subfigure}
        \caption{Execution time for attacking one target user in one repetition under the limited query syntax $\mathcal{Q}_{lim}$. We show the test accuracy on the (a) Census and (b) Insurance datasets using 5 repetitions, each repetition with 100 randomly selected users. We also show on the right the accuracy change of the average attack from one iteration to the next.}
        
    \label{fig:main-census-and-insurance}
\end{figure}

Figure~\ref{fig:main-census-and-insurance} shows the average test accuracy of the best attacks discovered by the fully-automated methods at every point of their running time on the Census and Insurance datasets. Both methods show similar performance as on the Adult dataset -- QuerySnout starts with a better-performing solution at the start, but QueryCheetah quickly catches up. QueryCheetah needs an order of magnitude less time than QuerySnout to terminate. At the end, the solutions discovered by QueryCheetah outperform the solutions discovered by QuerySnout. 

\section{Attacking all users}\label{sec:appendix_vulnerabiltiies}
\begin{figure}
    \centering
    \includegraphics[width=0.9\linewidth]{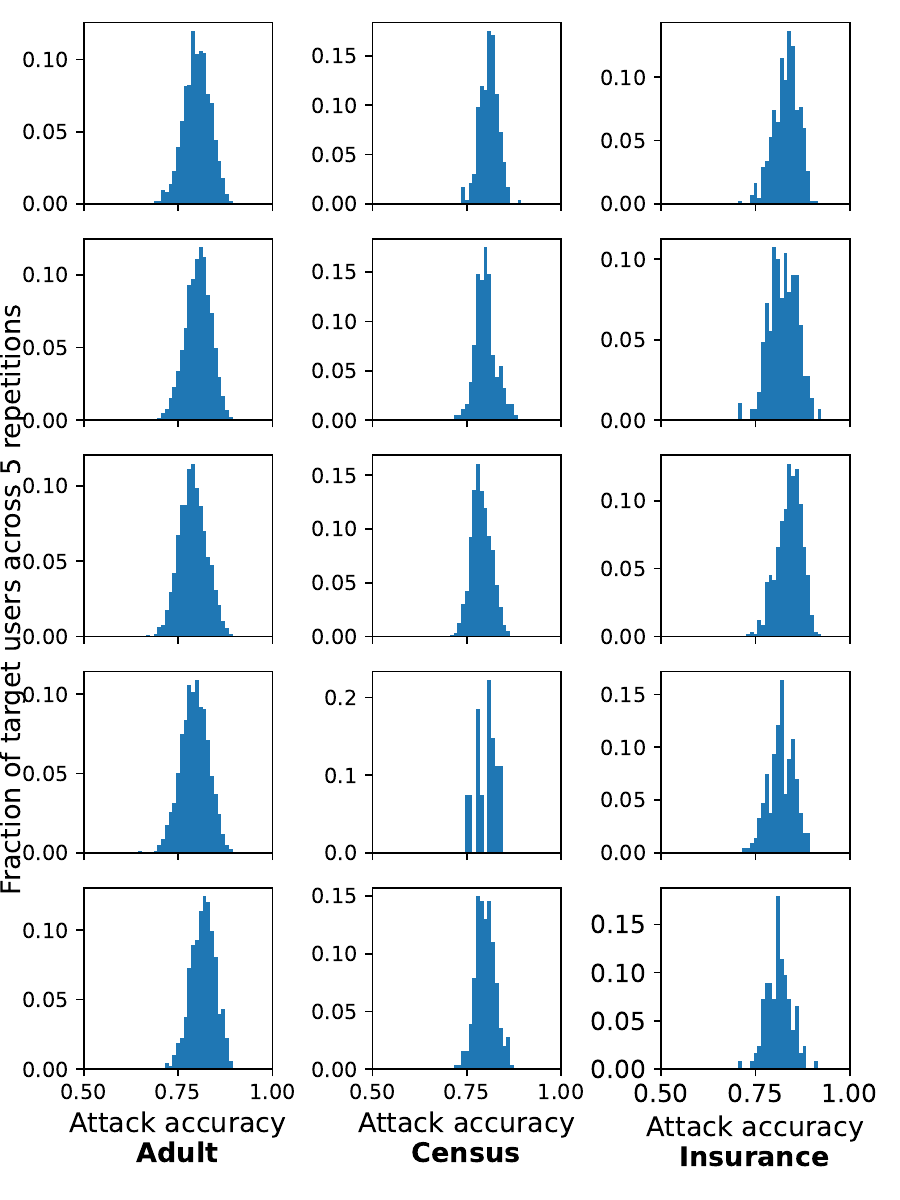}
    \caption{Vulnerabilities of all users in 5 repetitions on the 3 datasets.}    \label{fig:user_vulnerabilities_distribution_appendix}
\end{figure}

QueryCheetah's speed enables automated privacy auditing in reasonable time on all users across datasets, each with many repetitions. For each repetition on the three datasets considered in this paper, we have instantiated QueryCheetah on all unique users. Figure~\ref{fig:user_vulnerabilities_distribution_appendix} shows the discovered vulnerabilities.

\section{Type of attributes}
In this section, we explore if the type of selected dataset attribute impacts the performance of discovered attacks. More precisely, when sampling $n=5$ attributes, we sample $3$ ordinal and $2$ categorical, and instantiate QueryCheetah to find privacy vulnerabilities.

\subsection{Attacking the extended syntax} \label{sec:appendix_full_syntax}
In this setup,  QueryCheetah has discovered attacks for both Adult and Census in the extended syntax that outperform attacks in the limited syntax. The setup ensures that queries can include range conditions for 3 attributes, which might be a contributor factor to the performance jump.

\begin{table}
    \centering
    \begin{tabular}{lcc}
         & Adult & Census \\ \hline \hline
        \multirow{ 2}{*}{Limited syntax $\mathcal{Q}_{lim}$} & $80.26\% $ & $83.12\% $   \\
          & $\pm 0.45\%$ & $\pm 0.62\%$  \\ \hline
         \multirow{ 2}{*}{Extended syntax $\mathcal{Q}_{ext}$} & $81.91\%$ & $84.03\%$ \\
          & $\pm 0.55\%$ & $\pm 0.41\%$ \\
    \end{tabular}
    \caption{Accuracies of attacks discovered by QueryCheetah over 5 repetitions, each on 100 users, when there are 3 ordinal attributes.}
    \label{tab:extended_syntax_appendix}
\end{table}

\begin{table*}
    \centering
    \begin{tabular}{llrr}
        & & Adult  & Census  \\\hline
        \multirow{3}{*}{Limited syntax  $\mathcal{Q}_{lim}$} & \multirow{1}{*}{No mitigations} &  $80.26\% $ $\pm 0.45\%$ & $83.12\% $ $\pm 0.62\%$   \\
        \cline{2-4}
         & \multirow{1}{*}{Post-hoc mitigations} & $72.35\%$ $\pm 4.05\%$ & $64.10\%$ $\pm 4.43\%$  \\
         \cline{2-4}
         & Including mitigations in the search & $74.00\%$ $\pm 3.55\%$ & $66.03\%$ $\pm 4.58\%$  \\
         \hline 
         \multirow{3}{*}{Extended syntax  $\mathcal{Q}_{lim}$} &  \multirow{1}{*}{No mitigations} & $81.91\%$ $\pm 0.55\%$ & $84.03\%$ $\pm 0.41\%$ \\ 
         \cline{2-4}
         & \multirow{1}{*}{Post-hoc mitigations} & $72.05\%$ $\pm 4.13\%$ & $63.26\%$ $\pm 4.65\%$ \\ 
         \cline{2-4}
         & Including mitigations in the search & $75.89\%$ $\pm 3.87\%$ & $67.18\%$ $\pm 4.43\%$ \\
    \end{tabular}
    \caption{Impact of mitigation on attacks discovered by QueryCheetah over 5 repetitions, each on 100 users, when there are 3 ordinal attributes.}
    \label{tab:mitigations_appendix}
\end{table*}

\subsection{Mitigations} \label{sec:appendix_mitigations}
The mitigations are also effective in this setup with $3$ ordinal attributes.  Their post-hoc application to attacks leads to around $20\%$ drop in attack accuracy for Census. Finding workarounds these mitigations by searching for new attacks helps non-significantly.

\begin{table}
    \centering
    \begin{subtable}{0.48\textwidth}
    \begin{tabular}{lrr}
          & QueryCheetah & QuerySnout \\ \hline \hline
        Number of queries & $66.30 \pm 5.04$ & $26.88 $ $\pm 6.49$ \\
        Number of unique queries & $38.91 \pm 2.84$ & $24.15 $ $\pm 5.63$ \\
        Attack accuracy using & \multirow{2}{*}{$79.07\% \pm 0.90\%$} & \multirow{2}{*}{$75.23\% $ $\pm 1.48\%$} \\
        the subset of queries & & \\
        Percentage of accuracy & \multirow{2}{*}{$97.89\% \pm 0.48\%$} & \multirow{2}{*}{$96.70\% \pm 0.73\%$}  \\
        accounted for by the subset & & \\
    \end{tabular}
    \caption{Census dataset}
    \label{tab:attack_explainability_census_restricted_syntax}
\end{subtable}
\begin{subtable}{0.48\textwidth}
    \begin{tabular}{lrr}
          & QueryCheetah & QuerySnout \\ \hline \hline
        Number of queries & $68.46 \pm 4.47$ & $29.04 $ $\pm 6.30$ \\
        Number of unique queries & $45.86 \pm 3.64$ & $26.56 $ $\pm 5.78$ \\
        Attack accuracy using & \multirow{2}{*}{$81.88\% \pm 1.48\%$} & \multirow{2}{*}{$77.70\% $ $\pm 1.12\%$} \\
        the subset of queries & & \\
        Percentage of accuracy   & \multirow{2}{*}{$98.88\% \pm 0.55\%$} & \multirow{2}{*}{$96.96\% \pm 0.82\% $}  \\
        accounted for by the subset & & \\
    \end{tabular}
    \caption{Insurance dataset}
    \label{tab:attack_explainability_insurance_restricted_syntax}
    \end{subtable}
    \caption{Limited syntax $\mathcal{Q}_{lim}$: Performance when isolating all the difference-like queries from the discovered attacks on the (a) Census and (b) Insurance datasets. The comparison is performed over 5 repetitions on 100 users.}
\end{table}

\begin{table}
    \centering
    \begin{subtable}{0.48\textwidth}
    \begin{tabular}{lrr}
    
          &   & Generalized  \\
          & Difference-like  &  difference-like \\
          & queries &  queries \\ \hline \hline
        Number of queries & $24.23 \pm 5.53$ & $54.88 $ $\pm 7.50$ \\
        Number of unique queries & $15.02 \pm 3.69$ & $34.56$ $\pm 4.60$ \\
        Attack accuracy using  & \multirow{2}{*}{$67.59\% \pm 1.58\%$} & \multirow{2}{*}{$78.92\% $ $\pm 0.84\%$} \\
        the subset of queries & & \\
        Percentage of accuracy  & \multirow{2}{*}{$82.91\% \pm 2.06\%$} & \multirow{2}{*}{$96.78\% \pm 1.00\%$}  \\
        accounted for by the subset & & \\
    \end{tabular}
    \caption{Census dataset}
    \label{tab:attack_explainability_census_extended_syntax}
\end{subtable}
\begin{subtable}{0.48\textwidth}
    \begin{tabular}{lrr}
    
          &   & Generalized  \\
          & Difference-like  &  difference-like \\
          & queries &  queries \\ \hline \hline
        Number of queries & $17.59 \pm 3.01$ & $38.60 $ $\pm 6.96$ \\
        Number of queries & $12.08 \pm 2.05$ & $27.31 $ $\pm 4.62$ \\
        Attack accuracy using  & \multirow{2}{*}{$68.41\% \pm 1.47\%$} & \multirow{2}{*}{$77.47\% $ $\pm 1.98\%$} \\
        the subset of queries & & \\
        Percentage of accuracy  & \multirow{2}{*}{$80.21\% \pm 1.48\%$} & \multirow{2}{*}{$90.80\% \pm 1.98\% $}  \\
        accounted for by the subset & & \\
    \end{tabular}
    \caption{Insurance dataset}
    \label{tab:attack_explainability_insurance_extended_syntax}
\end{subtable}
\caption{Extended syntax $\mathcal{Q}_{ext}$: Performance when isolating all difference-like and generalized difference-like queries from the discovered attacks on the (a) Census and (b) Insurance dataset. The comparison is performed over 5 repetitions on 100 users.}
\end{table}

\section{Additional details about QuerySnout}\label{sec:addition_details_about_querysnout}
QuerySnout~\cite{cretu2022querysnout} maintains a population of multisets and iteratively improves it over $I$ iterations. First, it initializes the population $P_0$ by constructing multisets of randomly sampled queries. Namely, as $v_n=1$, the only degrees of freedom in $\mathcal{Q}_{lim}$ are the comparison operators, and thus, queries are sampled by sampling at uniform random the comparison operators $(c_1, \dots, c_n)$ from $\mathcal{C}^n=\{=, \neq, \perp\}^n$. Then, QuerySnout evaluates the fitness of all $P$ multisets in iteration $i-1, i\geq1$, $\{\bar{F}(S_{1,i-1}), \dots, \bar{F}(S_{P,i-1})\}$ and creates the population $P_{i}$ in iteration $i$ by (1) copying the top $P_e$ multisets with highest fitness values, $top\_k(P_{i-1}, \bar{F}, P_e):=argmax_{\{j_1, \dots, j_k\}}\sum_{j \in \{j_1, \dots, j_k\}}\bar{F}(S_{j,i-1})$ and (2) applying random modifications following hand-crafted rules to some of the remaining multisets. Finally, the attacker selects the multiset in $P_I$ with the highest fitness value and attacks the target QBS by sending the queries in it.

\section{Analysis of discovered attacks}\label{sec:appendix_analysis_of_attacks}
We analyze the discovered attacks by (1) isolating the subset of queries that have a given syntax, (2) evaluating their answers, and (3) fitting a logistic regression model, as per the AIA privacy game described in Section~\ref{subsec:game}, and (4) reporting the accuracy. We perform the four steps over 5 repetitions on 100 users and calculate the mean accuracy. To determine how well the subsets of a given type explain the attack accuracy, we calculate the ratio between the mean accuracy of the subset and the mean accuracy of the full attack over the same users and repetitions.

Our results show that difference-like queries and generalized difference-like queries account for most of the attack accuracy in the limited $\mathcal{Q}_{lim}$ and the extended syntax $\mathcal{Q}_{ext}$, respectively. Note that difference-like query pairs have the syntax of difference query pairs, shown in Equation~(\ref{eq:difference_queries}), but do not necessarily fulfill the uniqueness condition of difference queries. Similarly, generalized difference-like queries extend the syntax of difference-like queries to the extended syntax $\mathcal{Q}_{ext}$ and are only defined by their syntax. Thus, considering only their syntax is sufficient to identify difference-like and generalized difference-like queries.

For example, the attack discovered by QueryCheetah in the limited syntax for a user with non-sensitive attribute values ("occupation", "nativecountry", "hoursperweek", "race", "relationship") = (7,4,40,4,1), contains the following difference-like query pair:
\begin{gather}\label{eq:ex_difference_queries}
\begin{aligned}
    q_1 := \text{SELECT} & \; \text{count()} \;\text{FROM} \; D \\
    \text{WHERE} & \; \mathbf{nativecountry \; \neq \; 4} \; AND \; hoursperweek \; = \; 40  \\
    &  AND \; race \; = \; 4 \; AND \; sens \; \neq \; 1, \\
    q_2 := \text{SELECT} & \; \text{count()} \;\text{FROM} \; D \\
    \text{WHERE} & \; \hphantom{nativecountry \; \neq \; 4 \; AND} hoursperweek \; = \; 40 \\
    &  AND \; race \; = \; 4 \; AND \; sens \; \neq \; 1.
\end{aligned}
\end{gather}
The difference between $q_1$ and $q_2$ is bolded. Note that we map the values of the categorical attributes to integers $0, 1, \dots$, as described in Section~\ref{sec:datasets}. 
For the same user in the extended syntax $\mathcal{Q}_{ext}$, the attack discovered by QueryCheetah contains a generalized difference-like query pair that generalizes the query pair above:
\begin{gather}\label{eq:ex_generalized_difference_queries}
\begin{aligned}
    q_1 := \text{SELECT} & \; \text{count()} \;\text{FROM} \; D \\
    \text{WHERE} & \; \mathbf{nativecountry \; \neq \; 4} \; AND \\ & \; hoursperweek \; BETWEEN \; (40,41) \; AND \\
    &  race \; = \; 4 \; AND \; sens \; \neq \; 1, \\
    q_2 := \text{SELECT} & \; \text{count()} \;\text{FROM} \; D \\
    \text{WHERE} & \; hoursperweek \; BETWEEN \; (40, 41) \; AND \\
    &  AND \; race \; = \; 4 \; AND \; sens \; \neq \; 1.
\end{aligned}
\end{gather}

\section{Generalizability to budget-based QBSs}\label{sec:appendix_dp_qbss}
We here discuss possible strategies for extending QueryCheetah to budget-based QBSs such as QBSs implementing differential privacy guarantees. Recall that, by design, QueryCheetah only requires asking $m$ queries to the target QBS in order to attack a target user $u$. Indeed, as described in Sec.~\ref{subsec:game}, only the discovered best-performing multisets of $m$ queries are sent to the target QBSs, and the reported attack accuracies are measured. This is because the query multiset search is not performed on the target dataset and target QBS. The search is instead performed on datasets $D_1^{train}, \dots, D_f^{train}, D_1^{val}, \dots, D_g^{val}$ sampled by the attacker from an auxiliary dataset $D_{aux}$ and protected by QBSs instantiated using the executable software. All the queries used during the search are sent to these QBSs.

To target a budget-based QBS answering queries with privacy budget $\varepsilon$, the attacker needs to divide the budget between the $m$ queries of the discovered multiset. First, reducing the number of queries used, by using a smaller $m$, might help in this case to discover stronger attacks. Using fewer queries can be beneficial in this case as it limits the amount of noise added to each query~\cite{cretu2022querysnout}. Second, using an attack optimization proposed by QuerySnout~\cite{cretu2022querysnout}, a query $q_i$ with multiplicity $w_i$ in the multiset of $m$ total queries can be assigned a partial budget $\frac{w_i}{m}$ of the total privacy budget, $\frac{w_i}{m}\cdot\varepsilon$. Cretu et al.~\cite{cretu2022querysnout} formally showed that this attack optimization leads to more accurate query answers than averaging $w_i$ query answers each using a fraction $\frac{\varepsilon}{m}$ of the budget. Third, we identify two possible strategies for dividing the budget between the unique queries. The first strategy is to use the importance score $\bar{f}(q_{i})$ by assigning $\frac{w_i\cdot \bar{f}(q_{i})}{\sum_{j\in\{j_1,\dots,j_y\}}w_j\bar{f}(q_{j})}$ of the total budget to query $q_i$, where  $j_1,\dots,j_y$ denote the indexes of unique queries in the solution with multiplicities $w_{j_1},\ldots,w_{j_{y}}$, respectively. The second strategy is to learn a policy for dividing the budget between the queries as part of the search, which presents a line of future work.

\end{document}